\documentclass{aa}
\usepackage{graphicx}
\usepackage{txfonts}
\usepackage{booktabs}
\usepackage{natbib}
\usepackage{hyperref}
\hypersetup{colorlinks=true,linkcolor=blue,citecolor=blue}
\begin{document}

   \title{Characterizing NGC 6383: A study of pre-main sequence stars, mass segregation, and age using Gaia DR3 and 2MASS}

   \author{L.M Pulgar-Escobar
          \inst{1}\thanks{E-mail: \href{mailto:lescobar2019@udec.cl}{lescobar2019@udec.cl}}
          \and
        N.A Henríquez-Salgado\inst{1}
        \and 
        R.E. Mennickent\inst{1}
        \and
        Pierluigi Cerulo\inst{2}
          }

   \institute{Departamento de Astronomía, Universidad de Concepción, Casilla 160-C, Concepción, Chile
         \and
             Departamento de Ingeniería Informática y Ciencias de la Computación, Universidad de Concepción, Chile
             }

   \date{Received XXXX; accepted YYYY}

  \abstract
   {This study focuses on the young open cluster NGC 6383, situated in the Carina-Sagittarius arm within the Sh 2-012 star formation region. Previous studies have provided estimates of the cluster’s distance, age, and structural properties. The presence of pre-main-sequence and low-mass stars, combined with the new data from Gaia DR3 and with 2MASS, significantly enhances the relevance of studying this cluster.}
   {We aim to accurately identify cluster members, determine fundamental parameters, assess mass segregation, and establish precise age and distance using Gaia DR3 and 2MASS data.}
   {We employed Bayesian analysis and machine learning techniques, including the Hierarchical Density-Based Spatial Clustering of Applications with Noise (HDBSCAN) for member identification, the No-U-Turn Sampler (NUTS) from PyMC for modeling, the Sagitta neural network for the identification and age estimation of pre-main sequence stars, and ASteCA for isochrone fitting.}   
   {We identified 254 probable cluster members with a mode cluster age of $3.53^{+1.40}_{-1.00}$ Myr and a distance of $1.11\pm 0.06$ kpc. The core and tidal radius were determined to be $1.95 \pm 0.19$ and $40.4 \pm 14.3\,\mathrm{arcmin}$, respectively. The analysis revealed primordial mass segregation among binary stars, indicating that NGC 6383 is not fully relaxed. The color magnitude diagram (CMD) shows a well-defined main sequence and a population of pre-main sequence stars, suggesting recent star formation activity from approximately 1 to 6 Myr ago. The software used for this investigation is released with the paper.}
   {Our analysis provides updated parameters for NGC 6383, confirming relative recent star formation and mass segregation, and demonstrating the effectiveness of combining advanced computational techniques with traditional methods for studying stellar clusters.}


    \titlerunning{Bayesian characterization of NGC 6383}
    \authorrunning{L.M Pulgar-Escobar et al.}

   \maketitle

\section{Introduction}\label{S_intro}
The estimation of membership is a crucial step in the identification and characterization of star clusters. Various methods have been proposed in \cite{1930LicOB..14..154T, 1949ApJ...110..117S, 2018AA...610A..30A, 1978MNRAS.182..607F, 1989MNRAS.236..263P, 1991MNRAS.249...76B, 2005AA...438.1163K, 2007AA...462..157P, 1968ArA.....5....1L}, each focusing on certain properties and sometimes leading to consensus or discrepancies. Recently, the implementation of Bayesian analysis for membership identification provides a different method to compare or complement the ones mentioned.

NGC 6383\footnote{NGC 6383, also known as NGC 6374 in the New General Catalog \citep{2008hsf2.book..497R} and classified as Collinder 334 and Collinder 335 in the Collinder catalog, was initially misclassified in the original Collinder catalog \citep{1931AnLun...2....1C}.} is a young open cluster located in the Carina-Sagittarius arm, within the Sh 2-012 star formation region \citep{1959ApJS....4..257S}. It is part of the larger Sirius OB1 association, along with NGC 6530 and NGC 6531 \citep{2008hsf2.book..497R}. The galactic coordinates are $\ell = 355.68^{\circ}$ and $b = 0.05^{\circ}$ \citep{2017A&A...600A.106C}. The distance to the cluster has been estimated by various authors, ranging from an upper limit of $2.13~\mathrm{kpc}$ \citep{1930LicOB..14..154T} to lower limits of $0.760~\mathrm{kpc}$ \citep{1949ApJ...110..117S} and $0.840~\mathrm{kpc}$ \citep{2018AA...610A..30A}. The cluster spans an angular size of $20.0\,\textrm{arcmin}$ \citep{2013A&A...560A..76M} and is estimated to be between $1.70$ and $5.00\,\textrm{Myr}$ old \citep{1978MNRAS.182..607F,1989MNRAS.236..263P,1991MNRAS.249...76B,2005AA...438.1163K,2007AA...462..157P}, though \citet{1968ArA.....5....1L} estimated an age of $20.0\,\textrm{Myr}$. The reddening $E(B-V) = 0.320 \pm 0.020$ derived by \citet{2008hsf2.book..497R} aligns with estimates from previous authors \citep{1971AAS....4..241B,1978MNRAS.182..607F,1978MNRAS.184..661L,1985AA...151..391T,1989MNRAS.236..263P,1994RMxAA..29..141F,2007AA...462..157P}, but differs from \citet{2018AA...610A..30A}, who determined a reddening of $E(B-V) = 0.51 \pm 0.03$. This reddening is due to a surrounding large shell-structured ionized HII region with a radius of approximately $1.00\deg$.

HD 159176, a double-lined spectroscopic binary composed of O7V type stars located in the projected center of the cluster, is responsible for ionizing the HII region \citep{2016ApJ...832..211P}. The estimated age of HD 159176 is $2.30-2.80$\,\textrm{Myr} \citep{2010AA...511A..25R}. A careful analysis of both the ages of the cluster and its central star system suggests that the central star predates the cluster itself. Furthermore, \citet{1978MNRAS.182..607F} proposed that HD 159176 initiated star formation in the cluster's core and beyond.

In a study by \citet{2018AA...610A..30A}, it is hypothesized that if the stars numbered as \textit{Star No. 6} have begun to evolve and HD 159176 is a blue straggler, the age of the cluster should be between $6.00$ and $10.0\,\textrm{Myr}$. The cluster parameters exhibit a wide range of values, as a result of the fact that their estimates were obtained with diverse methodologies. The cluster hosts sources that have recently entered the main sequence track, while pre-main sequence stars (PMS) are also present \citep{2019MNRAS.484.5102K}, leading to considerable debate regarding the cluster's age. A summary of the historical results can be seen in Table \ref{tab:literature}.

This paper is structured as follows: Section \ref{sec:methodology} outlines the methodology, covering data acquisition and processing methods, including the use of the Gaia DR3 and 2MASS catalogs, and describes the membership determination using HDBSCAN and the Bayesian analysis for parameter estimation. Section \ref{sec:results} presents the results, including the determination of the cluster's age, distance, and structural properties, the analysis of mass segregation, and the color-magnitude diagram. It also discusses the findings in the context of previous studies, addressing peculiar stars and comparisons with historical data. Finally, Section \ref{sec:conclusion} concludes the study, summarizing the key results and their implications. The photometric system adopted includes the Gaia DR3 photometric bands ($G, G_{BP}, G_{RP}$) and the 2MASS bands ($J, H, K_s$).

\begin{figure}
	\centering
	\resizebox{\hsize}{!}{\includegraphics{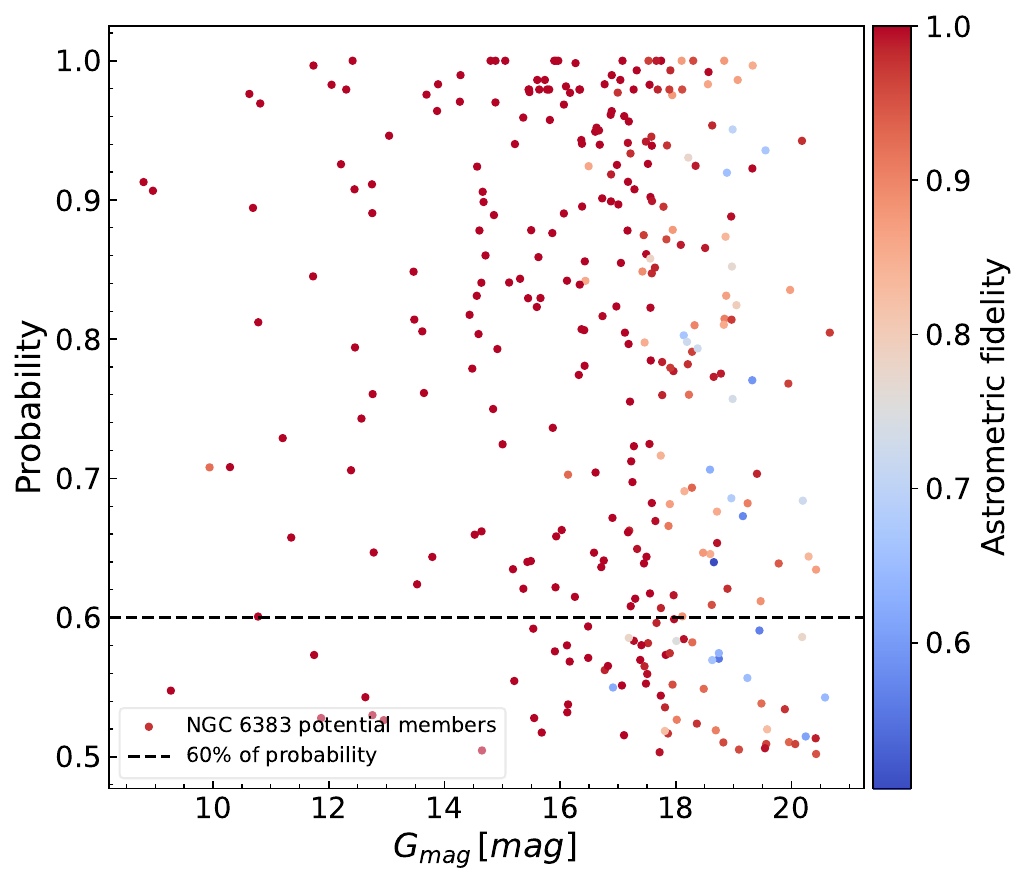}}
	\caption{Membership probability of potential NGC 6383 members (membership probability $> 50\%$) is plotted against their $G$ magnitudes, with colors representing astrometric fidelity. Each data point indicates a star, using a color gradient from blue to red to reflect the varying reliability of astrometric measurements. The horizontal dashed line at a probability of $0.6$ marks the threshold used to distinguish between probable members (with membership probabilities between $60\%$ and $80\%$) and members (with membership probabilities $\geq 80\%$) of the cluster. As shown in the figure, stars with fainter magnitudes tend to have lower astrometric fidelity. This becomes more evident at $G > 18~\mathrm{mag}$.}
	\label{fig:probabilities}
\end{figure}

\section{Methodology}\label{sec:methodology}
We acquired data from the \textit{Gaia} third Data Release (DR3) \citep{2016A&A...595A...1G,2023A&A...674A...1G}, executing a cone search of $40.0~\mathrm{arcmin}$ radius in the \textit{Gaia} archive, which yielded $23740$ sources. We restricted the initial selection of filtered sources to parallax ranges between $0.750$ and $1.10~\mathrm{mas}$-based on limits established in \citet{2024arXiv240301030P}.

To enhance data reliability, we applied the astrometric fidelity parameter from \cite{2022MNRAS.510.2597R}. This parameter, derived from a neural network analysis of 17 \textit{Gaia} catalog metrics, assesses the trustworthiness of the astrometric solution. Sources with an astrometric fidelity above 0.5 were retained, narrowing the selection to 15480 sources, which represent 65 percent of the initial data set.

\begin{figure}
	\centering
	\resizebox{\hsize}{!}{\includegraphics{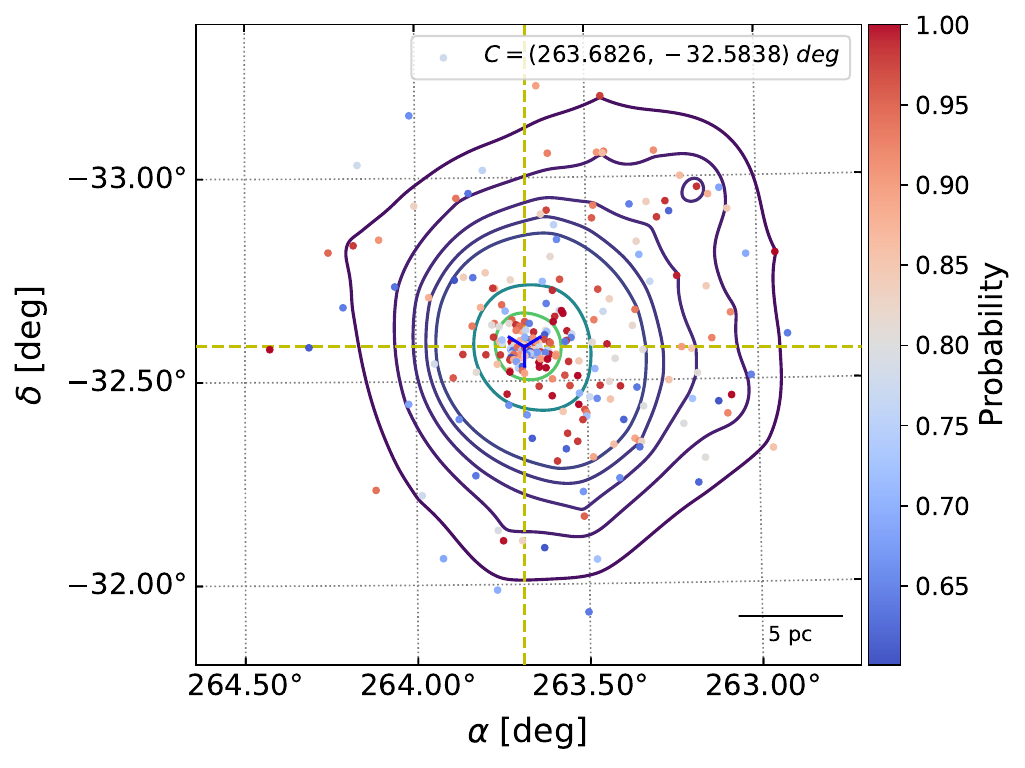}}
	\caption{Spatial distribution of probable members and members of NGC 6383 in Right Ascension $(\alpha)$ and Declination $(\delta)$, color-coded according to their membership probability. Each color indicates the probability of cluster membership. The contours represent levels of Kernel Density Estimation (KDE) with an exponential kernel, revealing the density gradient of the points. The central coordinates of the cluster are marked by white dashed lines at $ \alpha = 263.683^\circ $ and $ \delta = -32.584^\circ $, which correspond to the density peak within the plotted area. A scale bar of 5 parsecs is included at the bottom right to provide a reference for the spatial scale.}
	\label{fig:center}
\end{figure}

Subsequent refinement ensured the inclusion of only sources for which all parameters $(\alpha,\delta,\ell,b)$, proper motions $(\mu_{\alpha*},\mu_\delta)$, parallax ($\varpi$), and magnitudes $(G, G_\mathrm{BP}, G_\mathrm{RP})$ were available, reducing the dataset to $15276$ sources. Systematic parallax offsets identified in \citet{2021A&A...649A...2L} were corrected using the \href{https://pypi.org/project/gaiadr3-zeropoint/}{\textsc{Gaiadr3\_zeropoint}} package. Moreover, to address the bias in proper motion for bright sources ($\mathrm{G}<13~\mathrm{mag}$), as discussed in \citet{2021A&A...649A.124C}, we applied a magnitude-based correction for sources with $\mathrm{G} = 11-13\,\mathrm{mag}$, compensating for up to $80~\mathrm{\mu as\,yr^{-1}}$ discrepancy between the frames of reference for bright and faint sources.

Furthermore, for the infrared part of the spectrum, we obtained the $J$, $H$, and $K_s$ magnitudes from the \textsc{Two Micron All-Sky Survey} (2MASS) \citep{2006AJ....131.1163S}. A cross-match was performed to obtain a final list of sources, using the pre-computed crossmatch from Gaia DR3, as explained in \citet{2022gdr3.reptE..15M}. The table used for the crossmatch was \texttt{tmass\_psc\_xsc\_best\_neighbour}, giving $5333$ crossmatched sources with a separation distance of less than or equal to $0.3~\textrm{arcsec}$. The join type used was a left join, which is a method that maintains all the data from Gaia and adds the available 2MASS data for those sources, keeping the original $15276$  sources.

\begin{table*}
\caption{Historical review of observed parameters for the open cluster NGC 6383, gathered from various studies. The table details the number of cluster members, core and tidal radii, parallax, distance, E(B-V) color excess, distance modulus $(m - M)$, and estimated ages. This summary provides a view on how measurements and interpretations of NGC 6383 have evolved over time.}
\label{tab:literature}
\centering
\resizebox{\hsize}{!}{%
\begin{tabular}{llllllllllll}
\hline\hline
Reference & Members & Core Radius & Tidal Radius & Parallax & Distance & E(B-V) & $(m - M)$ & Age\\
& & arcmin & arcmin & mas & kpc & mag & mag & Myr\\
\hline
\citet{1930LicOB..14..154T} & $\cdots$ & $\cdots$ & $\cdots$ & $\cdots$ & 2.13 & $\cdots$ & $\cdots$ & $\cdots$\\
\citet{1949ApJ...110..117S} & $\cdots$ & $\cdots$ & $\cdots$ & $\cdots$ & 0.76 & $\cdots$ & $\cdots$ & $\cdots$\\
\citet{1961RGOB...27...61E} & $\cdots$ & $\cdots$ & $\cdots$ & $\cdots$ & $\cdots$ & 0.30 & $\cdots$ & $\cdots$\\
\citet{1967MNRAS.135..377G} & 74 & $\cdots$ & $\cdots$ & $\cdots$ & 1.38 & $\cdots$ & 10.7$\pm$0.5 & $\cdots$\\
\citet{1968PhDT........29E} & At least 21 & $\cdots$ & $\cdots$ & $\cdots$ & 1.30 & 0.35 & 10.6 & 5.0\\
\citet{1968ArA.....5....1L} & $\cdots$ & $\cdots$ & $\cdots$ & $\cdots$ & 1.25 & $\cdots$ & 10.5 & $\sim$ 20\\
\citet{1971AAS....4..241B} & $\cdots$ & $\cdots$ & $\cdots$ & $\cdots$ & 1.07 & 0.26 & 10.9 & $\cdots$\\
\citet{1978MNRAS.182..607F} & $\cdots$ & 1.25 & $\cdots$ & $\cdots$ & 1.50$\pm$0.20 & 0.33$\pm$0.02 & 10.9 & 1.7$\pm$0.4\\
\citet{1978MNRAS.184..661L} & $\cdots$ & $\cdots$ & $\cdots$ & $\cdots$ & 1.35 & 0.35 & 10.6 & Up to 5.0\\
\citet{1985AA...151..391T} & $\cdots$ & $\cdots$ & $\cdots$ & $\cdots$ & 1.40$\pm$0.15 & 0.30$\pm$0.01 & $\cdots$ & $\cdots$\\
\citet{1989MNRAS.236..263P} & $\cdots$ & $\cdots$ & $\cdots$ & $\cdots$ & $\cdots$ & 0.35 & 11.7 & $\sim$ 4.5\\
\citet{1991MNRAS.249...76B} & 27 & $\cdots$ & $\cdots$ & $\cdots$ & 1.38 & 0.34 & $\cdots$ & $\sim$ 20\\
\citet{1994RMxAA..29..141F} & $\cdots$ & $\cdots$ & $\cdots$ & $\cdots$ & 1.40 & 0.33 & $\cdots$ & $\cdots$\\
\citet{2003AA...407..925R} & $\cdots$ & $\cdots$ & $\cdots$ & $\cdots$ & $\cdots$ & 0.33 & 10.7 & $\cdots$\\
\citet{2005AA...438.1163K} & 13 & 4.80 & 15.0 & $\cdots$ & 0.99 & 0.30 & 10.9 & 5.0\\
\citet{2007AA...462..157P} & $\cdots$ & $\cdots$ & $\cdots$ & $\cdots$ & 1.70$\pm$0.30 & 0.29$\pm$0.05 & $\cdots$ & Less than 4.0\\
\citet{2008AA...477..165P} & $\cdots$ & $\cdots$ & 29.0$\pm$6.6 & $\cdots$ & 0.99 & 0.30 & 10.9 & 5.0\\
\citet{2010AA...511A..25R} & $\cdots$ & $\cdots$ & $\cdots$ & $\cdots$ & $\cdots$ & $\cdots$ & $\cdots$ & $\cdots$\\
\citet{2018AA...610A..30A} & $\cdots$ & $\cdots$ & $\cdots$ & $\cdots$ & 0.83$\pm$0.16 & 0.51$\pm$0.03 & 9.61$\pm$0.38 & $\sim$ 3-10\\
\citet{2019MNRAS.484.5102K} & At least 55 & $\cdots$ & $\cdots$ & $\cdots$ & $\cdots$ & $\cdots$ & $\cdots$ & 2.8$\pm$1.6\\
\citet{2021ApJ...923..129J} & 284 & $\cdots$ & $\cdots$ & 0.93$\pm$0.09 & 1.07 & $\cdots$ & $\cdots$ & $\cdots$\\
\citet{2022ApJS..262....7H} & $\cdots$ & $\cdots$ & $\cdots$ & 0.88$\pm$0.05 & $\cdots$ & 0.45 & $\cdots$ & 3.5\\
\citet{2024arXiv240305143H} & 322 & 2.49 & 22.7 & 0.89$\pm$0.08 & 1.10 & 0.38 & 10.2 & $\sim$ 4.0\\
\citet{2024arXiv240301030P} & 266 & 1.21$\pm$0.13 & 29.7$\pm$7.7 & 0.92$\pm$0.06 & 1.10$\pm$0.04 & 0.47$\pm$0.03 & 10.9$\pm$0.33 & $\sim$ 1-4\\
\hline
\end{tabular}}
\end{table*}
\begin{figure}
	\centering
	\resizebox{\hsize}{!}{\includegraphics{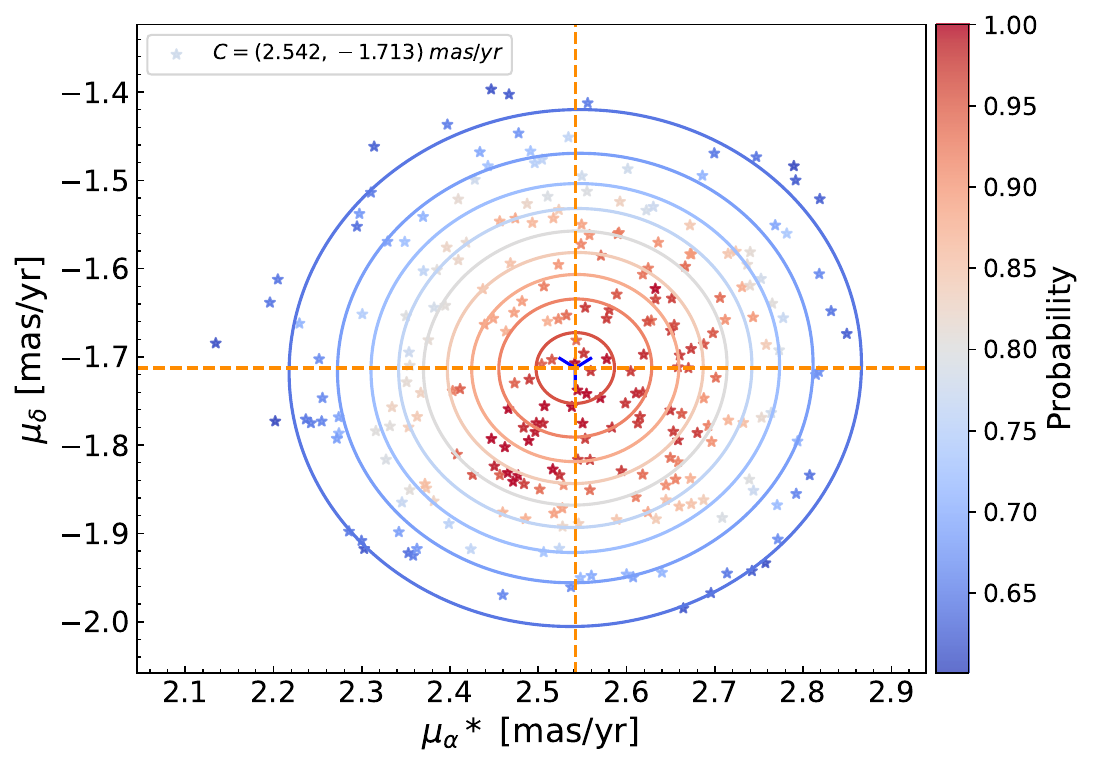}}
	\caption{Proper motions of probable members and members of NGC 6383 in Right Ascension ($\mu_{\alpha}^*$) and Declination ($\mu_{\delta}$), the symbols are color-coded according to the membership probability of the stars. The diagram uses hues transitioning from blue (lower probabilities) to red (higher probabilities), overlaid with density contours of proper motions. The contours illustrate the density gradient of the proper motions. The center of proper motion distribution is marked with a blue cross, indicating the inferred mean proper motion values $(\mu_{\alpha}^*, \mu_{\delta}) = (2.54, -1.71)~\mathrm{mas}\,\mathrm{yr^{-1}}$, with corresponding orange dashed lines highlighting these central coordinates.}
	\label{fig:proper_motion}
\end{figure}

\begin{figure*}
	\centering
	\resizebox{\hsize}{!}{\includegraphics{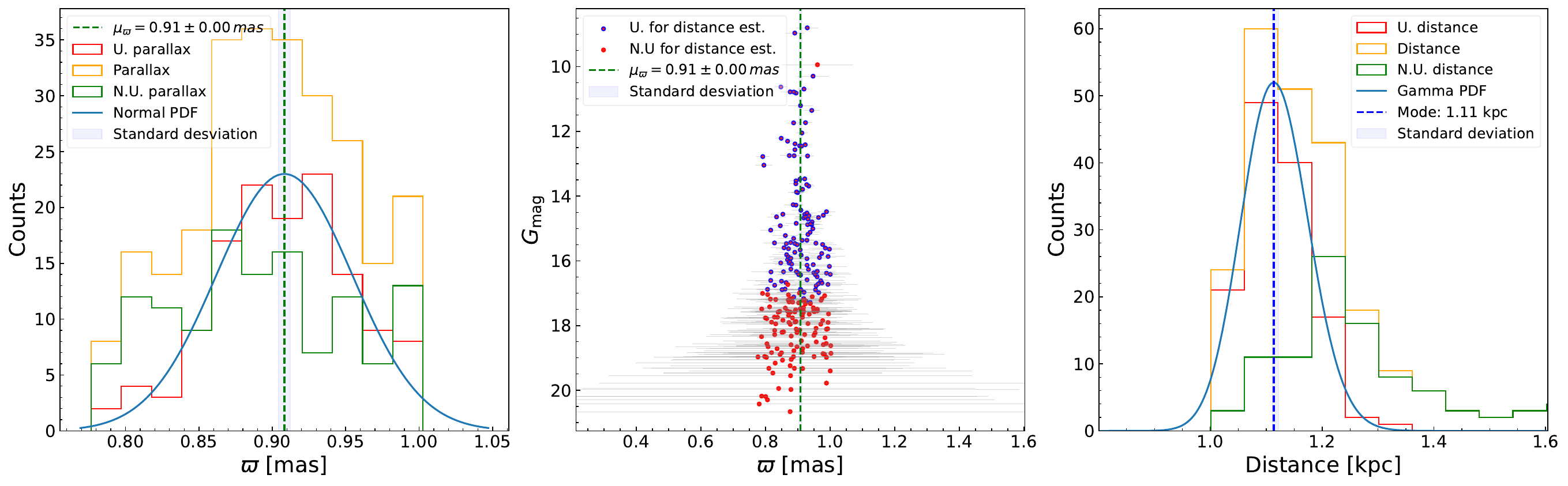}}
	\caption{\emph{Left panel:} Distributions of the parallax measurements in NGC 6383. \textit{U. parallax} in red represents measurements with fractional parallax errors less than $0.1$, used for the parallax estimation, \textit{Parallax} in orange represents all the data, and \textit{N.U. Parallax} in green represents sources with a fractional parallax error greater than 0.100, not used for the parallax estimation. The blue curve shows a Gaussian fit to the \textit{U. parallax} distribution, with the mean parallax value $ \mu_{\varpi} = 0.908 \pm 0.004 \, \mathrm{mas}$ marked with a dashed green line. \emph{Middle panel:} G magnitude versus parallax values, with the sources utilized for the distance estimation in blue and discarded due to having a fractional parallax error greater than 0.100 in red. Gray error bars illustrate the uncertainty in parallax measurements, especially pronounced at fainter magnitudes.  \emph{Right panel:} Histograms of geometric distances derived from \citet{2021AJ....161..147B} are shown, distinguishing between used (red) and all measured (orange) parallaxes, including those not used (green). The central dashed blue line indicates the mode of the sampled distance ($1.1100\, \mathrm{kpc}$), with its mean standard deviation of $0.0599\, \mathrm{kpc}$ shaded in blue. For convenience, only the portion of the histogram corresponding to distances within $0.8$ to $1.2$ times the minimum and maximum used distances is displayed.}
    \label{fig:parallax_distance}
\end{figure*}

\subsection{\sc COSMIC}
The Characterization Of Star clusters using Machine learning Inference and Clustering (\textsc{COSMIC}) developed by Lucas Pulgar-Escobar et al. (in prep.), is a suite of functions designed for analyzing open clusters. COSMIC utilizes unsupervised machine learning algorithms to process extensive datasets, such as those from \textit{Gaia}, to identify fundamental parameters of open clusters through clustering techniques and Bayesian estimation. As an open-source program, it is developed in \textsc{Python} 3.12 and integrates \textsc{PyMC} 5\footnote{\href{www.pymc.io}{www.pymc.io}}, a \textsc{Python} library specialized in Bayesian analysis.

The PyMC library aims to modernize and simplify Bayesian modeling. It offers a user-friendly interface for setting prior distributions and executes the sampling process efficiently. One of the key advantages of using PyMC is its implementation of the No-U-Turn Sampler (NUTS), an advanced MCMC algorithm that allows for efficient exploration of high-dimensional parameter spaces.

The NUTS algorithm builds upon Hamiltonian Monte Carlo (HMC) techniques, simulating a physical system to generate proposals. These proposals are based on the topology of the posterior distribution, allowing for an adaptive and balanced exploration of the parameter space. The algorithm terminates when a U-turn is detected, which serves as an indicator that the space has been sufficiently explored. A U-turn in the trajectory of the MCMC algorithm is detected when the trajectory of the sample path starts to bend on itself, indicating that the sampler is proceeding in the opposite direction.

Its efficiency and adaptability give NUTS an edge over traditional methods like Metropolis-Hastings and Gibbs sampling, which are often constrained by the shape of the distribution or the dimensions being considered \citep{phan2019composable}.

\subsubsection{Membership determination}\label{subsubsec:membership}

To identify potential members of NGC 6383, we employed the Hierarchical Density-Based Spatial Clustering of Applications with Noise (\textsc{HDBSCAN}) algorithm \citep{10.1007/978-3-642-37456-2_14,McInnes2017}, known for its effectiveness in identifying clusters of varying shapes, densities, and sizes without requiring a predetermined number of clusters \citep{2021A&A...646A.104H}, which makes \textsc{HDBSCAN} an ideal choice for this analysis, offering significant benefits over traditional clustering algorithms, especially in terms of robustness against noisy data and outliers.

After narrowing the data to a final subset of $15276$ sources, \textsc{HDBSCAN} was applied, focusing on proper motions as the primary clustering parameters.

To refine our selection of hyperparameters for optimal cluster identification, we utilized a systematic approach. This involved the iterative testing of various minimum cluster sizes to find the setup that enhances cluster size while ensuring high cluster separation strength, as indicated by the maximum $\lambda$ value. Our strategy aimed at selecting leaf clusters to generate numerous small, homogeneous groups, with the goal of including the maximum number of stars without compromising the separation strength. While this approach risks increasing false positives, it is essential for capturing all potential cluster sources, particularly for studying tidal sources. To mitigate this risk, we derived a pseudo-probability metric. This metric was calculated by determining the number of times a source was identified as a cluster member across multiple iterations, divided by the total number of iterations, and then multiplying this value by the HDBSCAN membership probability Sources with a resulting probability greater than $0.5$ were retained. We consider sources with a membership probability between 0.6 and 0.8 as probable members, while those with a probability greater than 0.8 are classified as members. The variation in cluster sizes and the threshold for minimum cluster size are plotted in Fig. \ref{fig:min_cluster_size}. Additionally, we used a Euclidean metric for clustering based on proper motions, finding that while the Haversine metric was tested, it produced non-physical clusters.

Finally, we used the \textsc{AstroPy Sigma Clipping} \citep{astropy:2013,astropy:2018,astropy:2022} utility to apply a $2\sigma$ outlier rejection around the mode of the parallax distribution in order to obtain the final sample of likely members of the cluster. This step provided the dataset of cluster members.
\begin{figure}
	\centering
	\resizebox{\hsize}{!}{\includegraphics{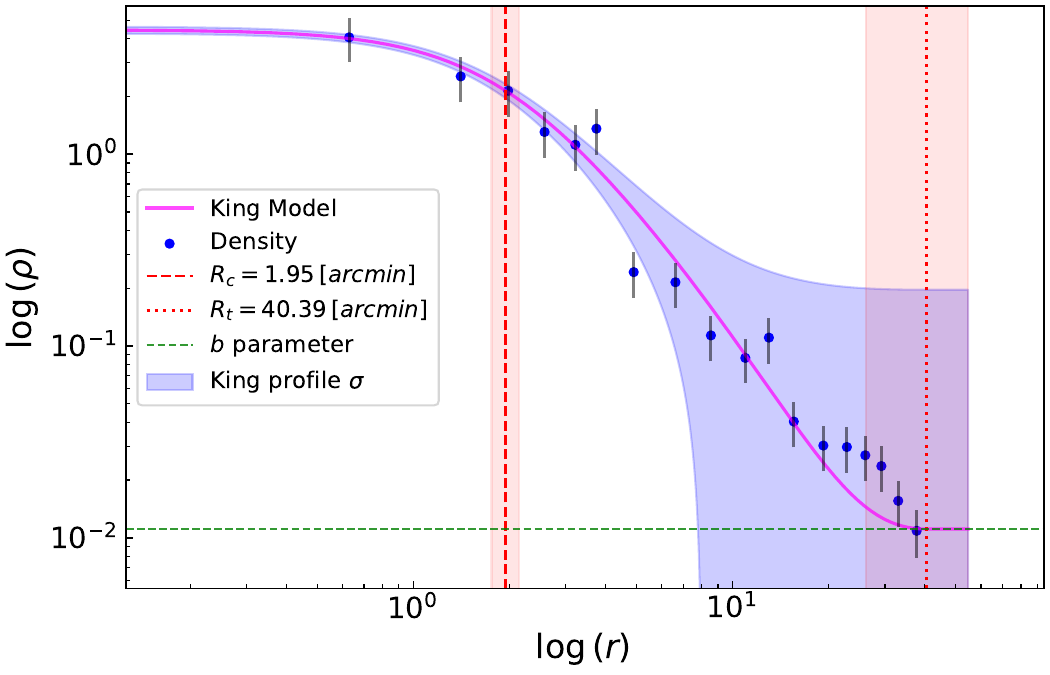}}
	\caption{Radial density profile of NGC 6383 shown in logarithmic scale. The plot overlays the observed stellar density (blue points) against the King model (solid fuchsia line). Dashed red and dotted red lines mark the core radius ($ R_c = 1.95 \, \text{arcmin} $) and the tidal radius ($ R_t = 40.4 \, \text{arcmin} $), respectively. The dash-dot green line represents the $ b $ parameter of the background level. The shaded area in blue indicates the $ 1\sigma $ uncertainty range of the King profile.}
	\label{fig:king}
\end{figure}

\begin{figure*}
	\centering
	\resizebox{\hsize}{!}{\includegraphics{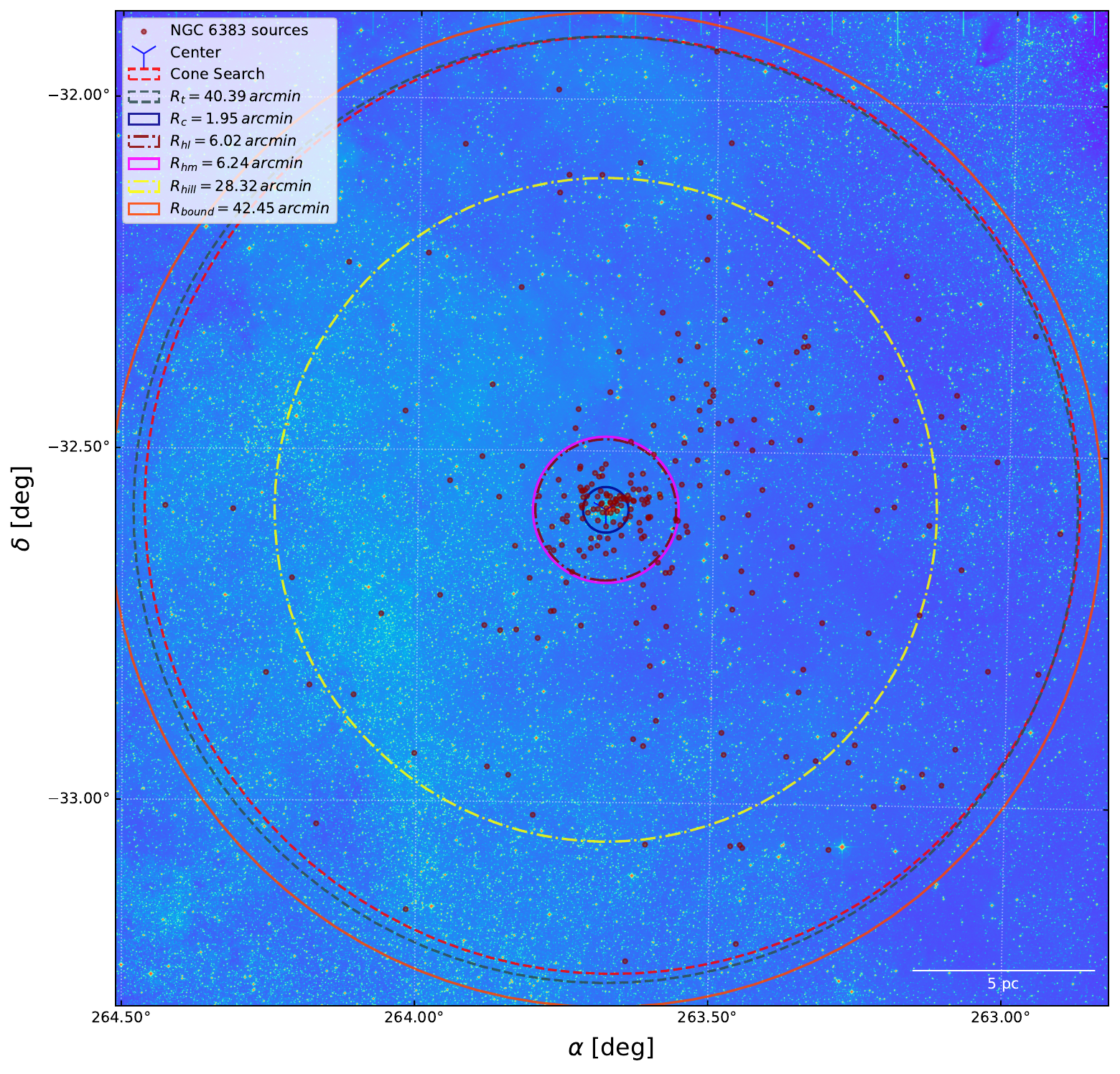}}
	\caption{Spatial distribution of probable members and members overlaid on a DSS2-red image, centered on the derived cluster center marked by a blue cross. The image details various structural and dynamical boundaries of the cluster as follows: the tidal radius $ R_t $ is depicted with a dashed dark slate gray line, the core radius $ R_c $ with a solid dark blue line, the half-light radius $ R_{\mathrm{hl}} $ with a dash-dot burgundy line, the half-mass radius $ R_{\mathrm{hm}} $ with a solid magenta line, the Hill radius $ R_{\mathrm{hill}} $ with a dashed yellow line, and the gravitational bound radius $ R_{\mathrm{bound}}$ with a solid orange line. A scale bar indicating $5\,\mathrm{pc}$ is positioned at the bottom right to provide spatial reference relative to the cluster scale.}
	\label{fig:real_sky}
\end{figure*}
\subsubsection{Parallax and Distance Estimation}
Distance estimation from parallax is challenging due to measurement errors, making the conversion from parallax to distance via the simple inverse method unreliable \citep{2015PASP..127..994B,2021AJ....161..147B}. Choosing an appropriate prior distribution is crucial. When analyzing the parallax measurements of stars within an open cluster, and assuming a simplified one-dimensional perspective, we assume that the parallax values follow a normal distribution. Additionally, we assume a gamma distribution for the distribution of distances.

To address this challenge, we obtained photo-geometric and geometric distances from \citet{2021AJ....161..147B}, who introduced a Bayesian approach incorporating a prior distribution tailored to different regions of the sky, improving distance estimates from Gaia EDR3 data. Geometric distances are derived using only parallax and prior information, while photogeometric distances combine parallax with photometric data to refine the distance estimate. However, we opted not to use photogeometric distances due to the presence of PMS stars and gas in the cluster, which could distort the magnitudes and colors.

In our hierarchical model, we calculate the mean $(\mu_{\mathrm{prior}})$ between the frequentist average distance derived from parallaxes and the distances, using $\mathcal{U}(0.5\mu_{\mathrm{prior}},1.5\mu_{\mathrm{prior}})$ as the prior for the cluster's mean distance. Our analysis is limited to members with fractional parallax errors below $0.1$ to ensure accuracy.

\subsubsection{Proper motions}
We used a two-dimensional Gaussian to model the distribution of the proper motions of the cluster members. A normal prior distribution was assigned to the mean proper motions $\left[\overline{\mu_\alpha},\overline{\mu_\delta}\right]$, based on frequentist means and standard deviations. For the standard deviations, we used half-Normal priors with a width equal to the frequentist standard deviation of the proper motions, and the correlation coefficient between the proper motions was uniformly distributed over the interval $[-1, 1]$. Additionally, we estimated the mean projected velocity of the cluster, defined as the quadrature sum of the proper motion components.
\begin{table}
	\caption{Summary of derived parameters for NGC 6383, focusing on sources with at least a 60 percent membership probability. Key measurements include distances assessed through parallax and distance modulus, mean age and stellar formation age, metallicity, and total stellar membership. The table also includes structural parameters such as core and tidal radii, absorption, and galactocentric distance, complemented by dynamic properties like proper motions, radial velocities, concentration parameters, specific radius measurements, the number of identified young stellar objects (YSOs), the fraction of YSOs, and the minimum segregation time.}
	\centering
	\resizebox{\columnwidth}{!}{%
		\begin{tabular}{llc}
			\hline
			\textbf{Parameter}                      & \textbf{Value}      & \textbf{Unit}                      \\ \hline
			Distance $(\varpi)$                     & $1.11 \pm 0.04$   & $\mathrm{kpc}$                     \\
			Distance (D.M.)                         & $1.15_{-0.13}^{+0.15}$   & $\mathrm{kpc}$                     \\
			Distance modulus                        & $10.3 \pm 0.3$    & $\mathrm{mag}$                     \\
			Age                                     & $3.53^{+1.40}_{-1.00}$           & $\mathrm{Myr}$                     \\
			Stellar formation range                 & $1.58 - 6.31$             & $\mathrm{Myr}$                     \\
			Metallicity ($Z$)                       & $0.024 \pm 0.008$   & -                                  \\
			Parallax $(\varpi)$                     & $0.908 \pm 0.004$   & $\mathrm{mas}$                     \\
			Number of members                       & $254$               & $\mathrm{stars}$                   \\
			Absorption $(A_V)$                      & $1.24 \pm 0.26$   & $\mathrm{mag}$                     \\
			Galactocentric distance $R_\mathrm{GC}$ & $7.190 \pm 0.004$             & $\mathrm{kpc}$                     \\
			Core radius $(R_c)$                     & $1.95 \pm 0.19$  & $\mathrm{arcmin}$                  \\
			Background $(b)$                        & $0.011 \pm 0.006$  & $\mathrm{stars\,arcmin^{-2}}$      \\
			Tidal radius $(R_t)$                    & $40.4 \pm 14.3$ & $\mathrm{arcmin}$                  \\
			Center density $(k)$                    & $4.91 \pm 0.44$   & $\mathrm{stars\,arcmin^{-2}}$      \\
			Hill radius                             & $28.3 \pm 1.0$            & $\mathrm{arcmin}$                  \\
			Gravitational bound radius              & $42.8 \pm 1.6$            & $\mathrm{arcmin}$                  \\
			Cluster center R.A.                     & $263.683 \pm 0.112$ & $\mathrm{deg}$                     \\
			Cluster center Dec.                     & $-32.584 \pm 0.112$ & $\mathrm{deg}$                     \\
			Proper motion R.A.                      & $2.540 \pm 0.009$  & $\mathrm{mas\,yr^{-1}}$            \\
			Proper motion Dec.                      & $-1.710 \pm 0.009$ & $\mathrm{mas\,yr^{-1}}$            \\
			Half-mass relaxation time               & $13.6 \pm 3.9$             & $\mathrm{Myr}$                     \\
			Radial velocity                         & $-15.1 \pm 31.1$  & $\mathrm{km\,s^{-1}}$              \\
			Concentration parameter $(C)$           & $3.03$             & -                                  \\
			Young stellar objects (YSOs)            & $53$                & $\mathrm{stars}$                   \\
			$Y_{\text{frac}}$                       & $0.28^{+0.06}_{-0.06}$              & -                                  \\
			Minimum segregation time                & $1.80 \pm 0.42$             & $\mathrm{Myr}$                     \\ \hline
		\end{tabular}%
	}
	\label{tab:ngc6383_results}
\end{table}
\begin{figure}
	\centering
	\resizebox{\hsize}{!}{\includegraphics{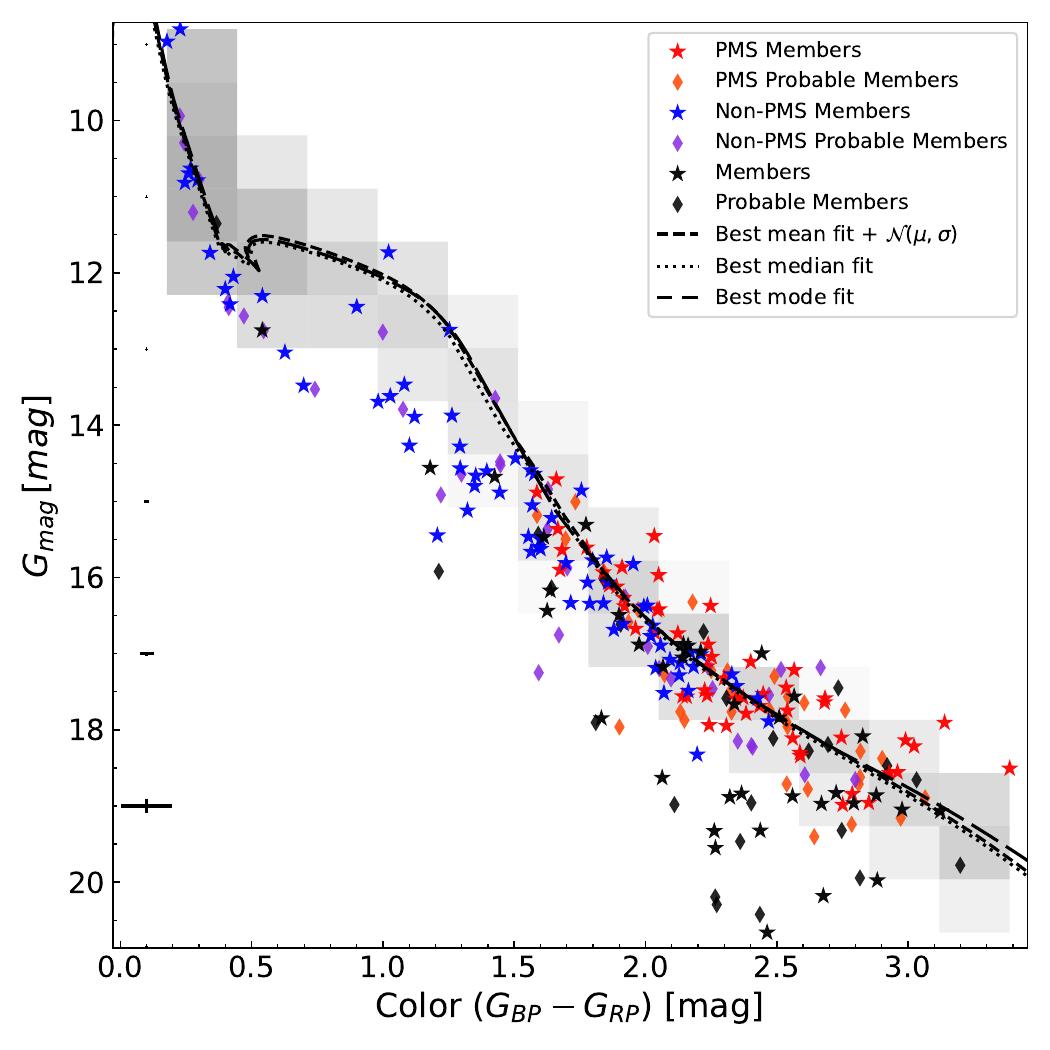}}
	\caption{CMD of NGC 6383 depicting the classification of cluster members and their evolutionary states, determined using Sagitta. The best-fit isochrones are represented by three lines: the dashed black curve for the mean fit, the dotted black line for the median fit, and the long dashed with offset line for the mode fit. Red stars indicate PMS Members with PMS probabilities $\geq 0.6$ and membership probabilities $\geq 0.8$. Orange diamonds represent PMS Probable Members with PMS probabilities $\geq 0.6$ and membership probabilities $< 0.8$. Blue stars show non-PMS Members, which are sources with PMS probabilities $< 0.6$ and membership probabilities $\geq 0.8$, indicating that they do not have enough probability to be classified as PMS, but they are not necessarily confirmed MS stars. Blue-violet diamonds represent non-PMS Probable Members, with PMS probabilities $< 0.6$ and membership probabilities $< 0.8$. Black stars and diamonds indicate Members and Probable Members with unavailable 2MASS data. The shaded area around the isochrone visualizes the uncertainty in the parameter fits, represented by a normal distribution. Additionally, error bars are plotted at regular intervals along the color axis to depict the median magnitude and color error.}
	\label{fig:cmd}
\end{figure}
\begin{figure*}
	\centering
	\resizebox{\hsize}{!}{\includegraphics{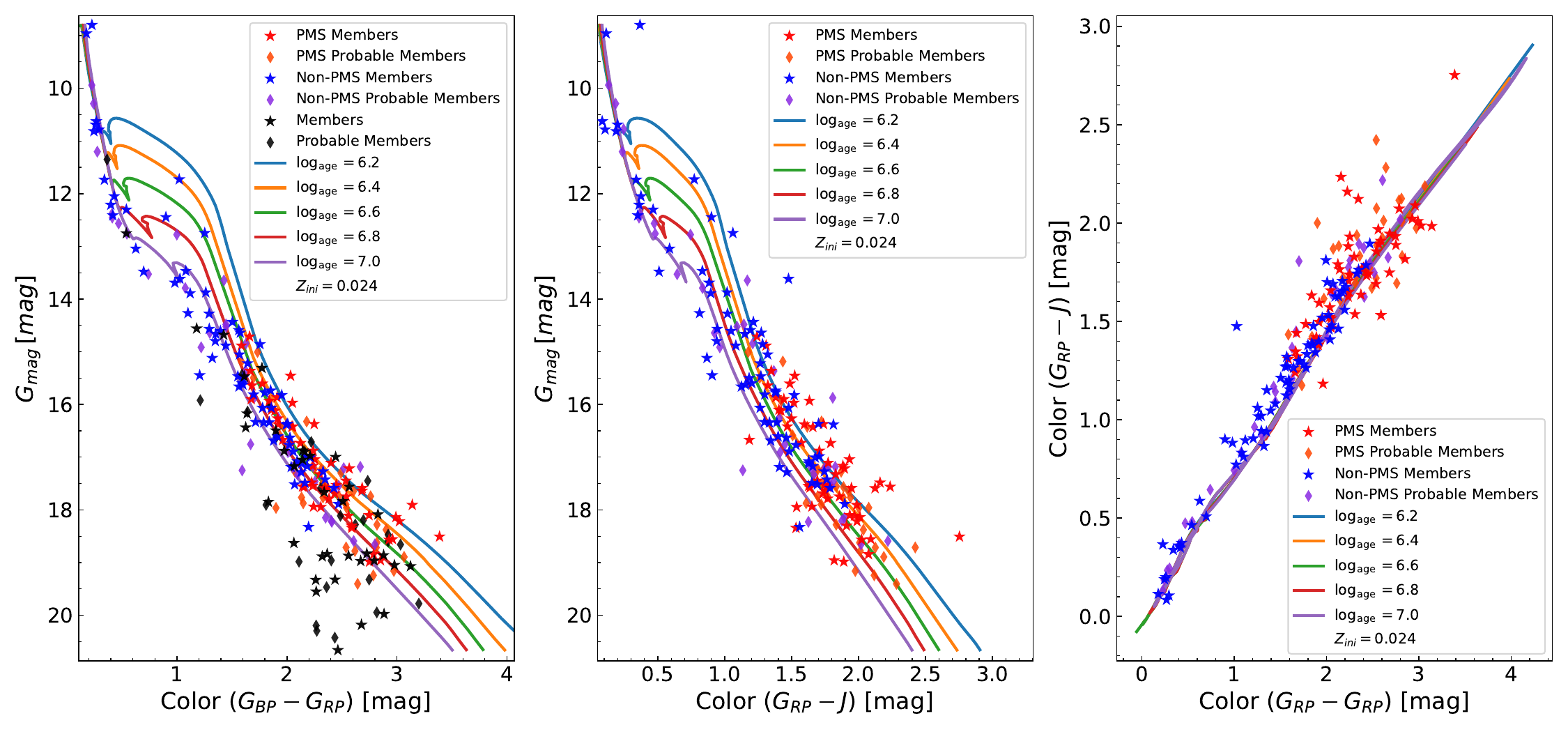}}
	\caption{\emph{Left panel:} CMD of NGC 6383 showing $G_{\text{mag}}$ against $G_{\text{BP}} - G_{\text{RP}}$ with isochrones for $\log(\mathrm{age}~\mathrm{yr}^{-1})$ ranging from $6.20$ to $7.00$. \emph{Middle panel:} CMD using $G_{\text{mag}}$ versus $G_{\text{RP}} - J$. \emph{Right panel:} Color-Color diagram $G_{\text{RP}} - G_{\text{RP}}$ against $G_{\text{RP}} - J$ with the equivalent color-color isochrones. The symbol notation follows that of Fig. \ref{fig:cmd}, representing different member classifications and evolutionary states within NGC 6383. Isochrones are color-coded to represent different ages, highlighting the evolutionary progression of members within NGC 6383. Each plot includes a legend indicating the initial metallicity ($Z_{\text{ini}} = 0.024$).
	}
 
	\label{fig:cmd_ngc6383_various}
\end{figure*}
\subsubsection{Center determination}
To determine the cluster's center, we used a weighted Kernel Density Estimation (KDE) from the Python library \textsc{sci-kit learn} \citep{scikit-learn}, assigning weights inversely proportional to the distance from the mean proper motion. Optimal KDE parameters were determined through Grid Search Cross-Validation within the package, exploring a range of bandwidths—from the mean positional error to the search cone radius—and all kernel types. The location of maximum density was taken as the cluster's center.

\subsubsection{Radial density profile}
To determine the structural parameters of the cluster, we utilized the \citet{1962AJ.....67..471K} density profile. The numerical density $\rho$, was computed by dividing the cluster area into concentric annuli, each containing an equal number of stars. The number of annuli, $K$, adheres to the equiprobable bin rule ($K = 2n^{2/5}$), where $n$ is the star count within the cluster. Fitting the King model to the density data allows for estimating the posterior distribution of the structural parameters.

Priors for the model parameters were set as follows: $b \sim \mathcal{U}(0, 2\rho_{\mathrm{min}})$, $k \sim \mathcal{U}(0, 2\rho_{\mathrm{max}})$, $R_c \sim \mathcal{U}(0, 0.8R_t)$, and $R_t \sim \mathcal{U}(R_c, 1.5T_{\mathrm{max}})$, where $T_{\mathrm{max}}$ represents the highest value between the Hill radius and the gravitational bound radius. This ensures that the chosen parameter space for $R_c$ and $R_t$ is within astrophysical valid limits.

The Hill Radius $(R_\mathrm{hill})$, which defines the region around a star cluster where its gravitational influence dominates over that of the Galaxy, can be calculated following \citet{2019A&A...627A.119C}:
\begin{equation}
R_\mathrm{hill} = R_{\mathrm{gc}} \times \left(\frac{m_c}{3M_{\mathrm{gc}}}\right)^{1/3}
\end{equation}
where $ R_{\mathrm{gc}} = 7.19 \pm 0.004\,\mathrm{kpc} $ denotes the galactocentric distance, $ m_c $ the cluster mass from \citet{2024arXiv240305143H}, with a value of $902.27 \pm 92.3~\mathrm{M_{\odot}}$, and $ M_{\mathrm{gc}} = 1.43 \times 10^{11} \pm 9.71 \times 10^7\,~\mathrm{M_{\odot}} $ the Galactic mass within the cluster's orbit, calculated as
\begin{equation}
	M_{\mathrm{gc}} = 2  \times 10^8\,M_\odot \left(\frac{R_{\mathrm{gc}}}{30\,\mathrm{pc}}\right)^{1.20}.
\end{equation}

Moreover, the gravitationally bound radius was estimated using the method described by \citet{1998MNRAS.299..955P}, which considers the cluster mass and Oort constants, using the formula
\begin{equation}
	r_\mathrm{bound} = \left( \frac{Gm_{\mathrm{c}}}{2(A-B)^2} \right)^{1/3},
\end{equation}
where $G$ is the gravitational constant, and $A$ and $B$ are the Oort constants, with values from \citet{2017MNRAS.468L..63B}.

\subsection{ASteCA}
Automated Stellar Cluster Analysis (ASteCA) \citep{2015A&A...576A...6P}, is an open-source Python tool designed to automate standard tests for determining basic parameters of stellar clusters. For accurate estimation of a cluster's metallicity, age, and extinction values, we employed ASteCA's isochrone fitting process. We utilized the \textsc{MIST} isochrones \citep{2016ApJS..222....8D, 2016ApJ...823..102C} along with the photometric systems of \textsc{Gaia} EDR3 and \textsc{2MASS}. The Initial Mass Function (IMF) settings followed \citet{2014ApJ...796...75C}, with fixed values for $\alpha = 0.090$, $\beta = 0.940$, and differential reddening set to zero.

The priors for metallicity $Z$, logarithmic age, and absorption $A_V$ were defined as uniform distributions ranging from $[0.001, 0.045]$, $[6.00, 7.00]$, and $[0.5, 2]$, respectively, aligned with findings from \cite{2024arXiv240301030P}. The distance modulus was modeled as $\mathcal{N}(10.3, 0.2)$, with the mean value corresponding to the expected distance modulus obtained from the parallax. We chose to use the mode of the posterior distribution for our analysis and utilized the NUTS from PyMC, prioritizing its efficiency over the Approximate Bayesian Computation (ABC) suggested by ASteCA.

\begin{figure}
	\centering
	\resizebox{\hsize}{!}{\includegraphics{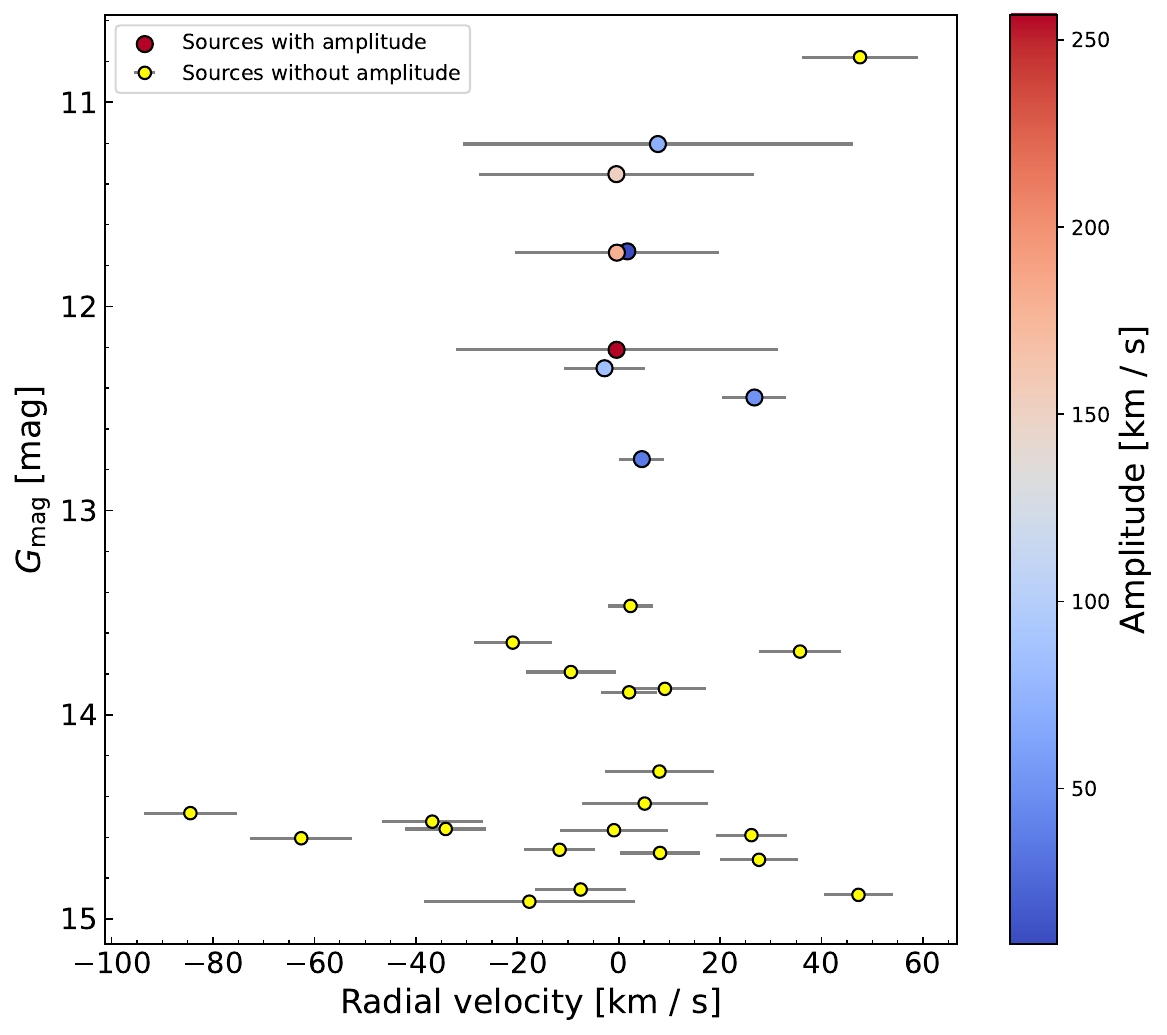}}
	\caption{Radial velocity plotted against G-band magnitude for NGC 6383 sources categorized by the presence of robust radial velocity amplitude data from Gaia. This measure reflects the total amplitude in the radial velocity time series following outlier removal. Yellow circles indicate sources without amplitude measurements, displaying radial velocity uncertainties with error bars. Blue circles represent sources with amplitude data, color-coded by amplitude value according to the scale on the right.}
	\label{fig:radial_velocities}
\end{figure}
\begin{figure}
	\centering
	\resizebox{\hsize}{!}{\includegraphics{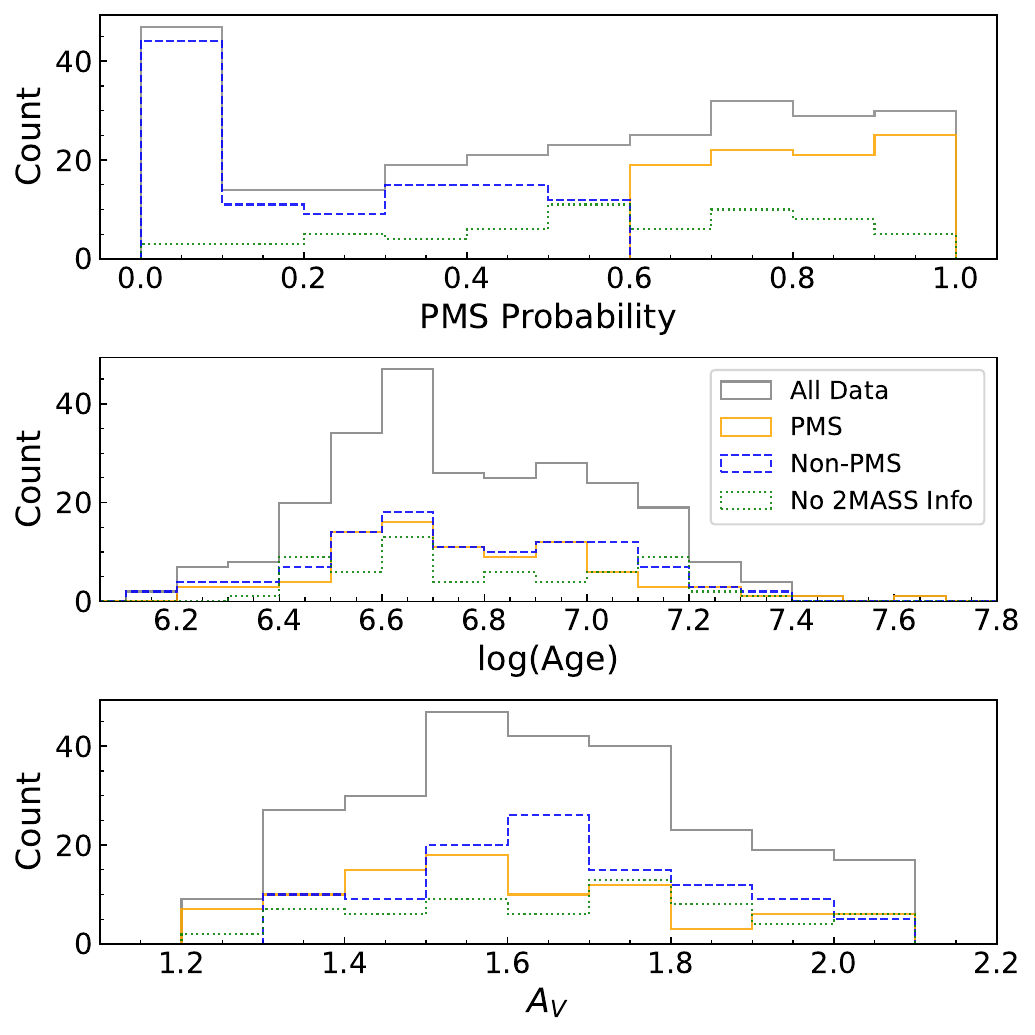}}
	\caption{Histograms of the parameters obtained with Sagitta for NGC 6383 probable members, including all data, members with a PMS probability over $60\%$, members with a PMS probability under $60\%$, and stars with missing 2MASS data. The upper panel shows the distribution of PMS probability values, indicating the likelihood of stars being PMS. The middle panel presents the distribution of logarithmic age values, illustrating the range and frequency of stellar ages within the cluster. The lower panel displays the distribution of visual extinction values ($A_V$), representing the amount of dust extinction affecting the observed stars.}
	\label{fig:pms_sagitta_stats}
\end{figure}
\subsection{Sagitta}
Sagitta is a neural network designed to classify stars as PMS and estimate their ages. It is trained with data from Gaia DR2 and the 2MASS. It incorporates three convolutional neural networks each dedicated to determining stellar extinction ($A_v$), PMS probability, and estimating stellar ages. Sagitta's age estimation is applicable for young stars up to $\sim 80~\mathrm{Myr}$ and relies on inputs including Gaia parallaxes, and average line-of-sight extinction \citep{2021AJ....162..282M}.

The tool’s efficacy stems from its training on a well-curated dataset of PMS stars within moving groups \citep{2020AJ....160..279K}, which is supplemented by various literature sources with previously measured ages. Sagitta uniquely offers stability in age predictions across G, K, and M-type stars within the same population, surpassing traditional isochrone techniques in consistency \citep{2023AJ....165..182K}.

The identification of PMS stars and the estimation of their ages within NGC 6383, allows for a detailed understanding of the cluster's star formation history to be gathered. It is important to note that Sagitta is most reliable for PMS stars, as the photometry of higher-mass stars that have reached the main sequence is not indicative of their age, limiting the model's applicability to stars that remain in the PMS stage.

\section{Results}\label{sec:results}
In this section, we report the results of our study, focusing on the mode of the distributions for members with a membership probability of at least $60$ percent.

Table \ref{tab:ngc6383_results} summarizes the parameters derived for sources with membership probabilities exceeding $60$ percent. The probability distribution for these sources is illustrated in Fig. \ref{fig:probabilities}. The condensed cluster tree is depicted in Fig. \ref{fig:condensed_cluster_tree}, and Fig. \ref{fig:center} indicates the cluster's center. The proper motions of the cluster members are shown in Fig. \ref{fig:proper_motion}, while Fig. \ref{fig:parallax_distance} shows the distance and parallax estimations. The structural parameters of the cluster, alongside the fitted King Profile, are presented in Fig. \ref{fig:king}. Fig. \ref{fig:real_sky} provides a visualization of the tidal radius, core radius, half-light radius, half-mass radius, the Hill radius, and the gravitational bound radius. Finally, Fig. \ref{fig:cmd} displays the best-fit isochrone and the CMD of NGC 6383.

Applying the HDBSCAN algorithm returned several clusters. A major flaw of HDBSCAN, as shown in \citet{2021A&A...646A.104H}, is its high false positive rate. This is due to the algorithm’s overconfidence, often reporting dense random fluctuations of a given dataset as clusters. For the problem of the determination of the membership, the proper motion of the stars was used to assess the significance of an identified cluster. As expected for the case of NGC6383, one significant cluster was identified. This cluster initially contained 701 sources, and after applying the $2\sigma$ clipping described in Section \ref{subsubsec:membership}, 321 sources were retained.

\begin{figure}
	\centering
	\resizebox{\hsize}{!}{\includegraphics{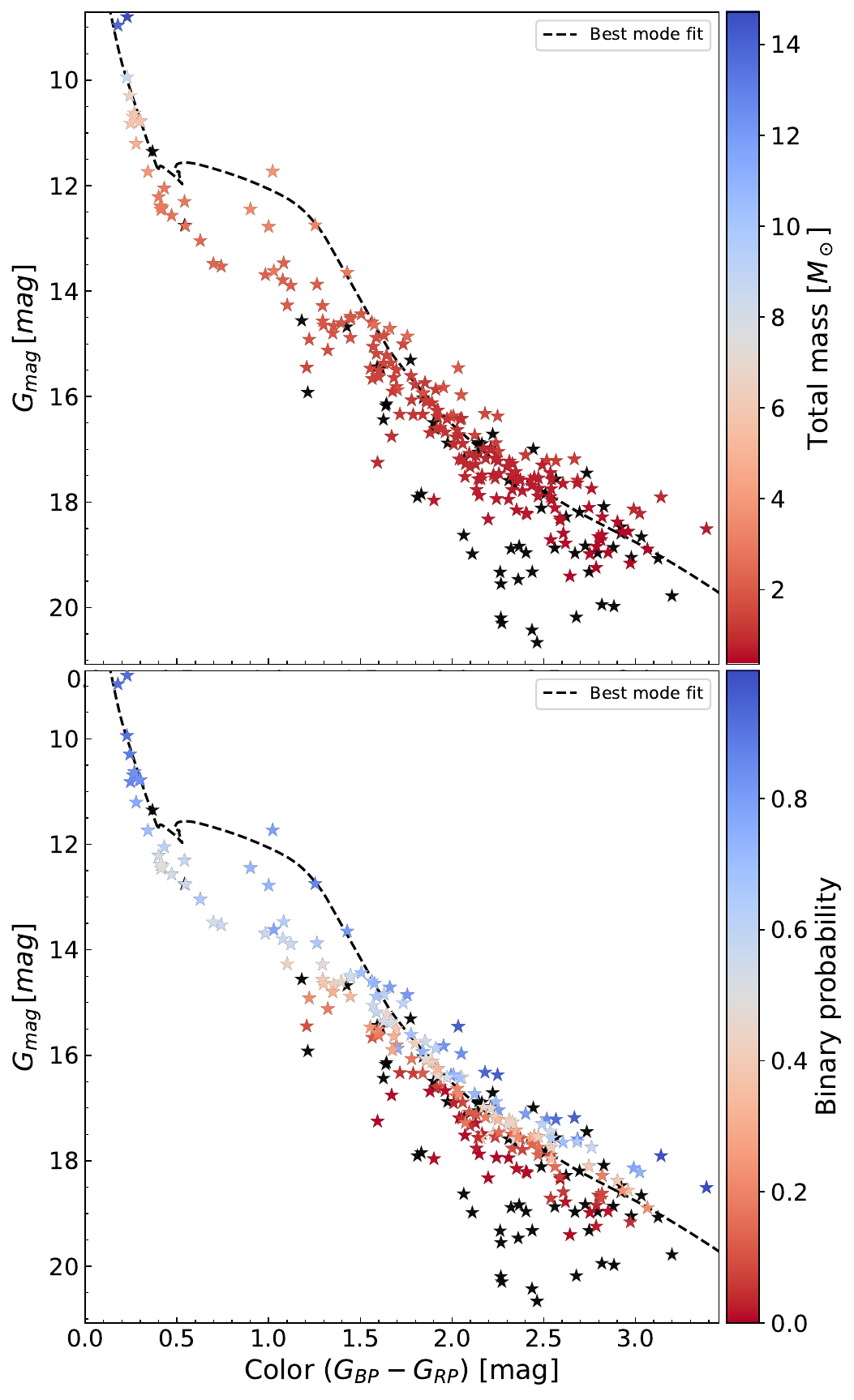}}
	\caption{\emph{Upper Panel:} CMD displaying probable members and members of NGC 6383, color-coded by total mass, ranging from lower (red) to higher mass (blue). The dashed black line represents the best mode fit isochrone, illustrating the evolutionary track inferred for the cluster. The total mass is calculated as the sum of the two components if the binary probability is greater than $0.7$. Black stars represent members and probable members with no available 2MASS data, and thus no calculated mass. \emph{Lower Panel:} The same stars as in the upper panel, now color-coded according to binary probability, illustrating the likelihood of binary systems within these stars, from low (red) to high (blue). The black dashed line again represents the best mode fit isochrone. Black stars represent members and probable members with no 2MASS data, thus lacking binary probability estimates.}
	\label{fig:cmd_mass_binary}
\end{figure}
\begin{figure}
	\centering
	\resizebox{\hsize}{!}{\includegraphics{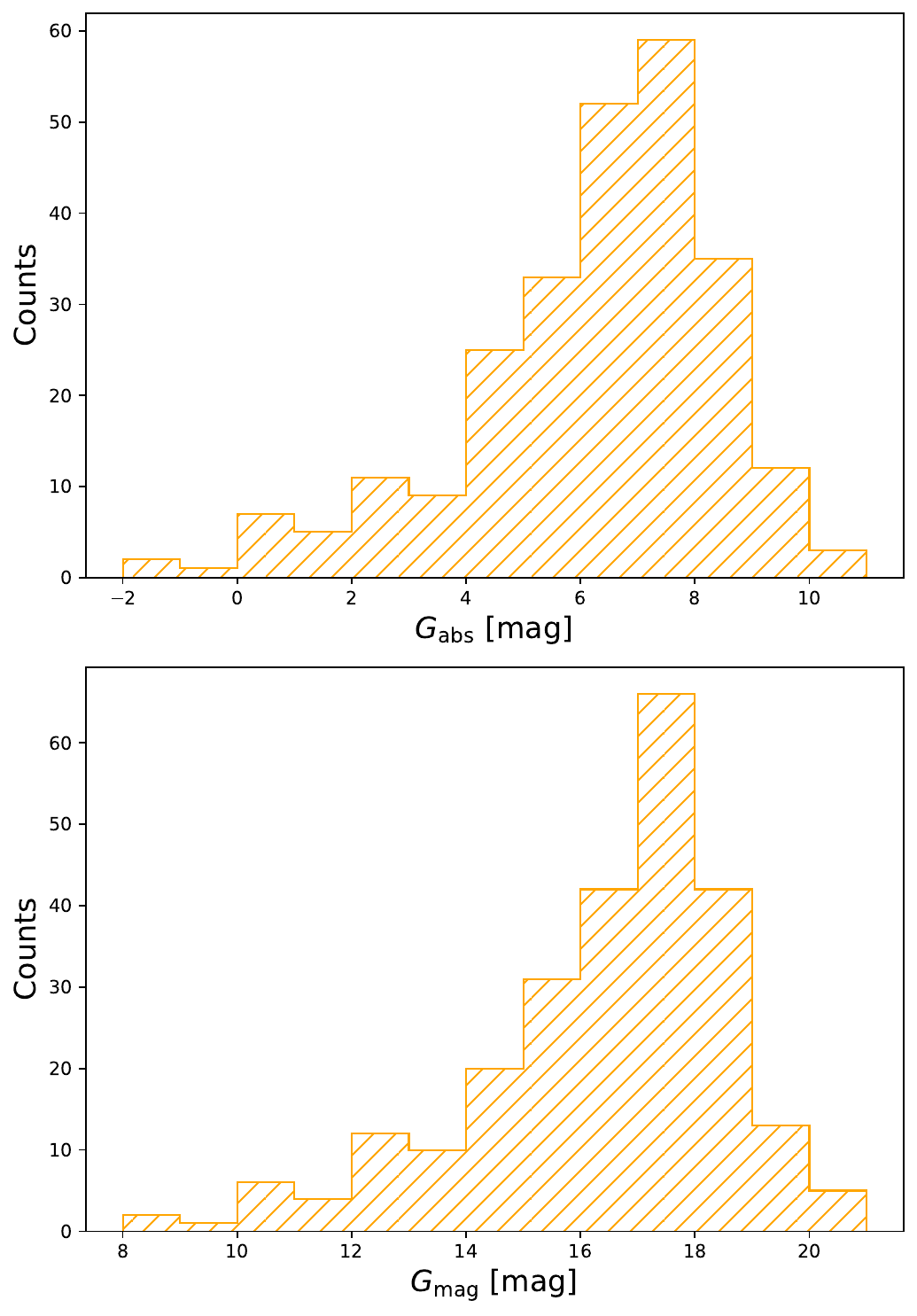}}
	\caption{\emph{Upper Panel:} Histogram representing the distribution of absolute magnitudes ($ G_{\text{abs}} $) for stars in NGC 6383 that have a membership probability of at least $60$ percent. \emph{Lower Panel:} Histogram of apparent magnitudes ($ G_{\text{mag}} $) for the same subset of stars, illustrating the observed magnitudes.}
	\label{fig:lf}
\end{figure}
\begin{figure}
	\centering
	\resizebox{\hsize}{!}{\includegraphics{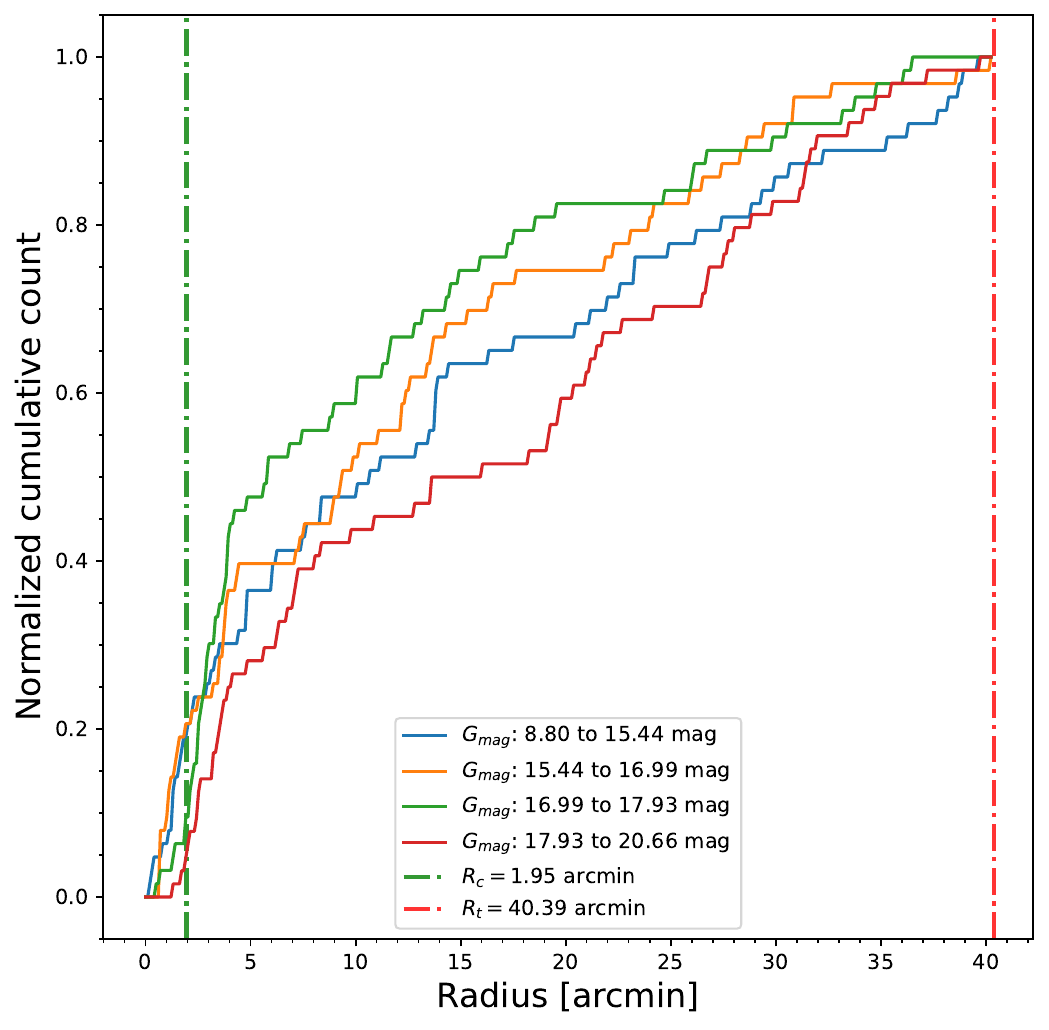}}
	\caption{Normalized cumulative counts of stars within NGC 6383 are plotted against radial distance from the cluster center for different brightness ranges, segmented into quartiles of $G_{\text{mag}}$ from the brightest at $8.80$ mag to the faintest at $\sim20.7$ mag. The green dashed line at $R_c = 1.95\,\mathrm{arcmin}$ marks the core radius, and the red dashed line at $R_t = 40.4\,\mathrm{arcmin}$ denotes the tidal radius, illustrating the spatial distribution of stars with varying brightness within these boundaries. See the text for more reference.}
	\label{fig:cumulative_luminosity}
\end{figure}
Regarding membership probabilities, NGC 6383 has $321$, $254$, $202$, and $161$ members with probabilities above $50$, $60$, $70$, and $80$ percent respectively. Correspondingly, $288$, $236$, $191$, and $153$ of these members are brighter than $G=19\,\mathrm{mag}$.

The central position of the cluster in equatorial coordinates is $263.6826 \pm 0.1122\,^\circ$ in RA and $-32.5838 \pm 0.1122\,^\circ$ in DEC. These errors are calculated from the quadratic sum of the mean errors in RA and DEC and the optimal KDE bandwidth. The mean proper motion values are $2.5400 \pm 0.0096\,\mathrm{mas\,yr^{-1}}$ in RA and $-1.7100 \pm 0.0086\,\mathrm{mas\,yr^{-1}}$ in DEC, with dispersions of $0.153$ and $0.138$ respectively, which align with the acceptable values from \citet{2023MNRAS.526.4107P}. The calculated projected velocity is $3.070 \pm 0.010\,\mathrm{mas\,yr^{-1}}$.

The mean parallax derived is $0.908 \pm 0.004\,\mathrm{mas}$, with a mode sampled distance of $1.1100 \pm 0.0600\,\mathrm{kpc}$. These values slightly deviate from the results obtained in earlier studies \citep{1930LicOB..14..154T,1949ApJ...110..117S,1978MNRAS.182..607F,1985AA...151..391T,2007AA...462..157P} but are consistent with more recent studies \citep{1971AAS....4..241B,2005AA...438.1163K,2008AA...477..165P,2021ApJ...923..129J,2022ApJS..262....7H,2024arXiv240301030P,2024arXiv240305143H}.

Radial velocities are estimated with a median, mean, and standard deviation of $-6.11$, $-15.1$, and $31.1 \,\mathrm{km}\,\mathrm{s}^{-1}$, respectively, using 16 clustered sources with probabilities above $60$ percent and a binary probability lower than $60$ percent. The substantial standard deviation primarily stems from the limited availability of radial velocity data, only $11.4$ percent of our Gaia-selected sources. The radial velocity calculation can be affected by sources with a relatively high radial velocity amplitude, which could indicate unresolved binaries and introduce significant variability in the measurements. To address this, as mentioned before, we only calculated the radial velocity of the cluster using sources with a binary probability lower than $60$ percent. This ensures that the derived cluster velocity is more representative of the overall motion of the cluster's stars, reducing the impact of outliers caused by binary systems. However, as mentioned, this value is still not fully representative of the cluster since the sample does not sufficiently capture the full dynamics of the cluster. Figure \ref{fig:radial_velocities} illustrates all the probable members and members with available radial velocities, their magnitudes, and the amplitude of their radial velocities.

\begin{figure*}
	\centering
	\resizebox{\hsize}{!}{\includegraphics{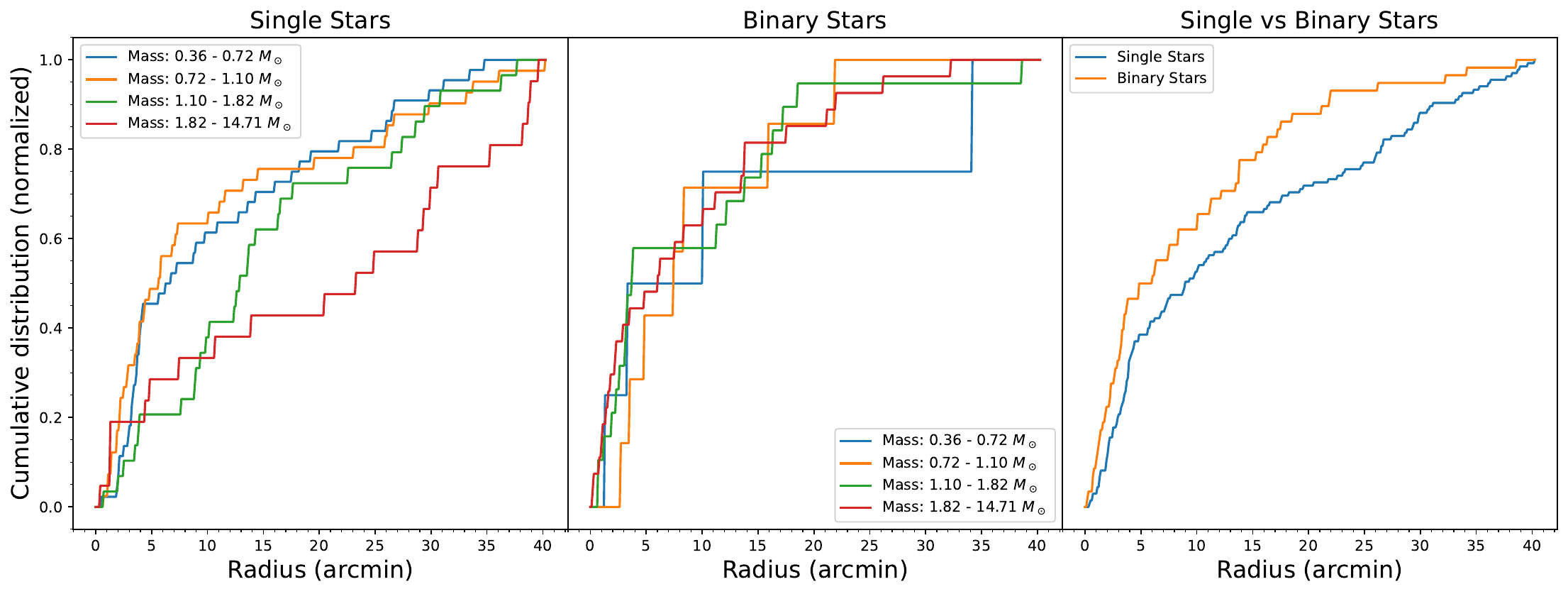}}
    \caption{\emph{Left panel:} The cumulative distributions of single stars within NGC 6383 are segmented into four mass quartiles ranging from $0.360 - 0.720\,M_\odot$ to $1.82 - 14.7\,M_\odot$, illustrating the spatial distribution across different mass segments and highlighting the potential influences of mass on the distribution of stars within the cluster. \emph{Middle panel:} The same distributions as in the left panel for binary stars only, where binaries are defined as sources with a binary probability greater than $60$ percent. \emph{Right panel:} The normalized cumulative distributions compare single stars to binary stars, with masses between $0.360\,M_\odot$ and $14.7\,M_\odot$. All plots are based on stars with a membership probability of at least $60$ percent. Kolmogorov-Smirnov (KS) test results are detailed in Sec. \ref{subsec:lum_mass_dyn}, to quantitatively assess the differences between distributions, underscoring the statistical validity of observed segregation patterns.}
	\label{fig:mass_seg}
\end{figure*}

\subsection{Structural Parameters}

The cluster exhibits a tidal radius of $40.4 \pm 14.3\,\mathrm{arcmin}$ ($13.10 \pm 4.71\,\mathrm{pc}$) and a core radius of $1.950 \pm 0.190\,\mathrm{arcmin}$ ($0.63200 \pm 0.00621\,\mathrm{pc}$). The King model parameters $k$ and $b$, representing structural constants, are $4.910 \pm 0.437$ and $0.01110 \pm 0.00630$ stars per square arcminute, respectively. The concentration parameter $C$, defined as $\log(R_t/R_c)$, is $3.03$, following \citet{1975AJ.....80..427P}. The half-light radius is determined to be $6.02000 \pm 0.00060\,\mathrm{arcmin}$ ($1.96000 \pm0.00020\,\mathrm{pc}$) and the half-mass radius is $6.240 \pm 0.251\,\mathrm{arcmin}$ $(1.650 \pm 0.066\,\mathrm{pc})$. Part of these results are represented in Fig. \ref{fig:king}.

The cluster’s outer limits extend to approximately $40.2\,\mathrm{arcmin}$ ($13.1\,\mathrm{pc}$), as shown in Fig. \ref{fig:center}. We identified $39$ sources between the Hill radius and the tidal radius, with membership probabilities ranging from $0.6$ to $1.0$, averaging $0.82$. These sources, plotted in Fig.\ref{fig:real_sky}, are likely not gravitationally bound to the cluster. This implies that they have proper motions similar to the cluster stars but are not necessarily physically associated with the cluster, due to the extensive area covered by the cone search.

\subsection{Age, Extinction, and Color-Magnitude Diagram}

The CMD of NGC 6383 is displayed in Fig. \ref{fig:cmd}, including PMS stars and the median, mean and mode isochrones. The mode indicates a logarithmic age in years of $6.550 \pm 0.145$, a distance modulus of $10.300\pm 0.262\,\mathrm{mag}$ corresponding to a distance of $1.150_{-0.130}^{+0.147}\,\mathrm{kpc}$, and a metallicity $Z $ of $0.024 \pm 0.008$. An estimated extinction correction of $A_V = 1.240 \pm 0.262$ was applied. The model’s distance modulus aligns with our parallax-derived estimates.

The CMD, particularly for sources brighter than $G=12.0\,\mathrm{mag}$, displays a well-defined main sequence. The presence of PMS stars and lower-mass main sequence stars broadens the lower part of the CMD, complicating the fit of a single isochrone. Figure \ref{fig:cmd_ngc6383_various} presents various isochrones for Gaia and 2MASS data, considering ages between $6.20$ and $7.00$ in logarithmic scale and initial metallicities $ Z $ of $0.024$. The age range of star formation in NGC 6383 appears to span from $6.20$ to $6.80$ in logarithmic scale.

Figure \ref{fig:cmd_ngc6383_various} also includes color-color diagrams $(G_\mathrm{BP} - G_\mathrm{RP})$ showing a clearer segregation of PMS stars, which tend to be at redder colors, as expected. Additionally, Fig. \ref{fig:pms_sagitta_stats} shows the returned Sagitta parameters, which include PMS probability, age, and $A_V$. The posterior distribution of the inferred parameters can be seen as a plot pair in Fig. \ref{fig:plot_pair_trace}.
\begin{figure*}
	\centering
	\resizebox{\hsize}{!}{\includegraphics{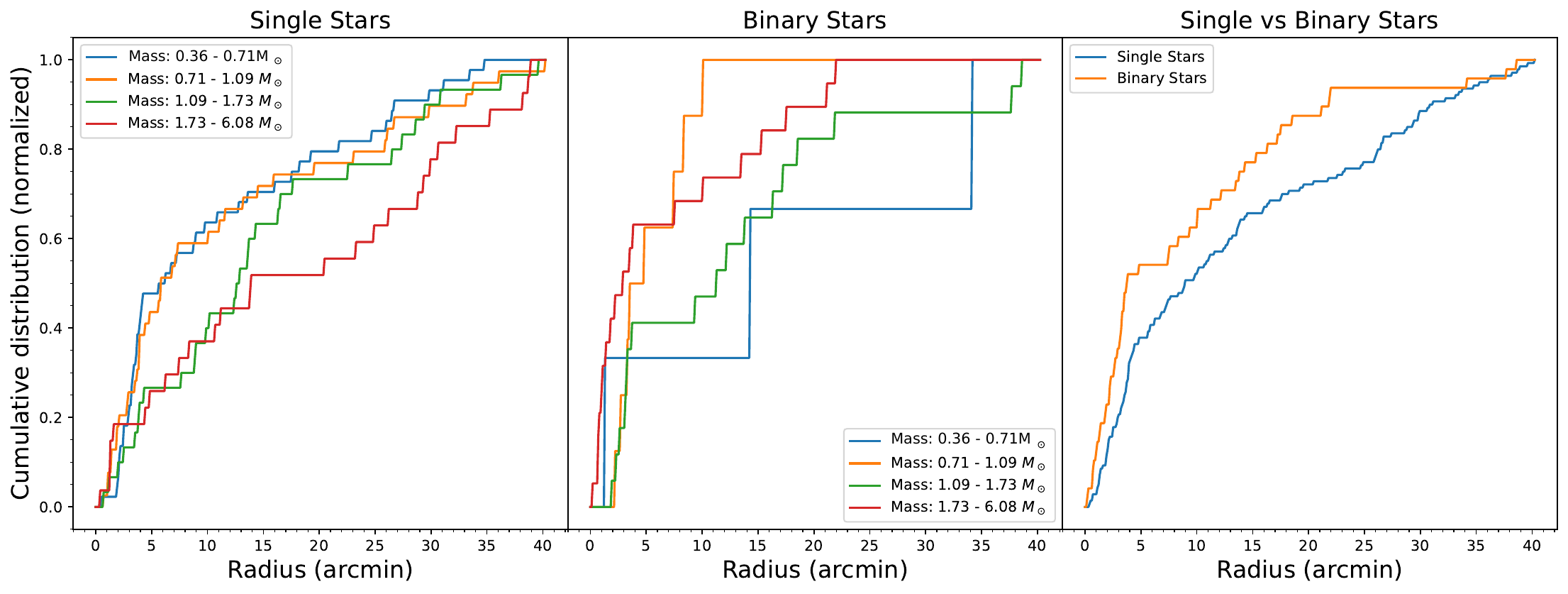}}
	\caption{\emph{Left panel:} Same as Fig. \ref{fig:mass_seg}, with cumulative distributions of single stars within NGC 6383. \emph{Middle panel:} The same distributions for binary stars. \emph{Right panel:} The normalized cumulative distributions compare single versus binary stars. These plots were generated with data filtered for a minimum segment mass cutoff of $6.08~M_\odot$, corresponding to a segregation time of $3.53~\mathrm{Myr}$. For a detailed discussion of the K-S test results, please refer to the main text.}
	\label{fig:mass_tseg}
\end{figure*}
\subsection{Identification of YSOs}

Young stellar Objects (YSOs) are early-stage stars still accreting material from their surrounding disks. They are characterized by significant variability due to the accretion processes. YSOs include both protostars and PMS stars. PMS stars, which encompass both classical T Tauri stars and Herbig Ae/Be stars, represent a later stage in stellar evolution. Thus, while PMS stars are a type of YSO, they are more evolved than protostars.

Our analysis aims to quantify the fraction of YSOs within the cluster, noted as $ Y_{\text{frac}} $, which serves as an age indicator. We determined the reddening-free parameter $ Q $ for each star using
\begin{equation}
	Q = (J - H) - \frac{E(J-H)}{E(H - K)} (H-K),
\end{equation}
where $ E(J-H)/E(H-K) $ is assumed to be $1.55$ \citep{1990ARA&A..28...37M}. Following \citet{2013MNRAS.436.1465B}, a star is considered a YSO if its $ Q $ value is less than $-0.050\,\mathrm{mag}$.

To estimate $ Y_{\text{frac}} $ more rigorously, we employed a Bayesian approach for determining confidence intervals for binomial proportions, which provides a more statistically robust estimate compared to traditional methods \citep{2011PASA...28..128C}. Using a non-informative $\mathcal{B}(1, 1)$ prior, the posterior distribution for $ Y_{\text{frac}} $ is modeled as
\begin{equation}
	Y_{\text{frac}} = \frac{N_{\text{YSO}}}{N_{\text{cl}}},
\end{equation}
where $ N_{\text{YSO}} $ is the number of YSOs and $ N_{\text{cl}} $ is the total number of cluster members. The $95\%$ credible interval for $ Y_{\text{frac}} $ was then derived from the posterior Beta distribution.

We identified 53 YSOs, with a mean $ Q $ value of $-0.520$, a median of $-0.380$, and a standard deviation of $0.420$ for sources below $-0.050\,\mathrm{mag}$.

The resulting $ Y_{\text{frac}} $ is $0.280$, with a $95\%$ credible interval of [$0.220$, $0.340$]. This value is higher compared to clusters analyzed in \citet{2013MNRAS.436.1465B}, where only 18 of 397 cluster candidates in their sample had $ Y_{\text{frac}} > 0.100 $, with no objects having $ Y_{\text{frac}} > 0.200 $. In contrast, our results indicates a higher fraction of YSOs compared to their sample. This difference suggests that NGC 6383 is relatively younger or has a higher recent star formation activity compared to most clusters in their study.

\subsection{Luminosity, Mass, and Dynamical State}\label{subsec:lum_mass_dyn}

The luminosity function (LF) represents the distribution of the magnitude of the cluster members. For NGC 6383, we converted apparent $G$ magnitudes to absolute magnitudes using the parallax-derived distance. Figure \ref{fig:lf} shows the LF, indicating an increase towards dimmer magnitudes, alongside the luminosity function derived from apparent magnitudes.

The radial cumulative distributions of cluster members at various magnitude levels, illustrated in Fig. \ref{fig:cumulative_luminosity}, show that fainter stars, ranging from $17.93$ to $20.66\,\mathrm{mag}$, are less concentrated around the cluster's center. This observation is consistent with Gaia's completeness limit for magnitudes between $18$ and $19$ at distances around $1\,\mathrm{kpc}$ \citep{2020MNRAS.497.4246B,2023AA...677A..37C}. The significance of this distribution was confirmed using a Kolmogorov-Smirnov (K-S) test, yielding p-values from $0.009$ to $0.052$ when compared with other ranges.

ASteCA calculates the probability of a star being part of a binary system and estimates the masses of the primary and secondary stars. However, these estimates are less reliable in the PMS stage due to the poorly defined CMD, as previously discussed. Figure \ref{fig:cmd_mass_binary} displays these binary probabilities and total mass, indicating that these values should be interpreted with caution, as they may not be entirely accurate. The average mass of the stars is $1.59\,M_\odot$. 

To investigate mass segregation, we constructed radial normalized cumulative mass distributions for single stars, binary stars, and the combination of the two types of stellar systems, as shown in Fig. \ref{fig:mass_seg}. The K-S test was employed to compare these distributions, indicating a significant difference between binary and single stars, with a p-value of $0.07$ and a K-S test statistic of $0.2079$. This suggests binary mass segregation, where binary stars are more centrally concentrated. Additionally, the last quartile mass range between $1.82$ and $14.71~M_\odot$ shows a significant difference compared to the other mass ranges, with p-values ranging from $0.04$ to $0.95$ and a K-S test statistic ranging from $0.14$ to $0.37$. However, this mass range is broad and cannot ensure mass segregation within it.

To contextualize these findings, we calculated the half-mass relaxation time $t_{\mathrm{rh}}$ \citep{1969ApJ...158L.139S}, which is crucial for estimating the minimum mass segregation time $t_{\mathrm{seg}}$. The half-mass relaxation time is given by:

\begin{equation}
t_{\mathrm{rh}} = \frac{0.17N}{\ln(\lambda N)}\sqrt{\frac{r_{hm}^3}{GM}}
\end{equation}

where $N$ is the number of cluster members, $\lambda = 0.110$ is a constant \citep{1994MNRAS.268..257G}, and $r_\mathrm{hm} = 1.650 \pm 0.066\,\mathrm{pc}$ is the half-mass radius, derived from the radial cumulative mass distribution. The calculated half-mass relaxation time for NGC 6383 is $13.60\pm 3.87\,\mathrm{Myr}$.

Using the half-mass relaxation time, we can estimate $t_{\mathrm{seg}}$ for the cluster's most massive star by following the approach of \citet{1969ApJ...158L.139S}:

\begin{equation}
t_{\mathrm{seg}} = \frac{\langle m \rangle}{m} t_{\mathrm{rh}}
\end{equation}

where $\langle m \rangle$ is the average mass of a star in the cluster, and $m$ is the mass of the most massive star. This results in a segregation time of $1.470 \pm 0.418\,\mathrm{Myr}$ for the most massive star of $14.7\,M_\odot$, indicating it has had sufficient time to undergo mass segregation.

We evaluated the potential for primordial mass segregation by comparing the radial distributions of stars with a $t_{\mathrm{seg}}$ greater than $3.53\,\mathrm{Myr}$, which corresponds to the age of the cluster. The sample includes stars with masses ranging from $0.360\,M_\odot$ to $6.08\,M_\odot$, which corresponds to $97.0$ percent of the original sample. The results, shown in Fig. \ref{fig:mass_tseg}, revealed a similar behavior to Fig. \ref{fig:mass_seg}, where the highest mass bin has a distinct distribution. This similarity is expected, given that the sample is nearly identical to the one used in the previous analysis. The mass range is narrower compared to the previous figure, with p-values ranging from $0.0600$ to $0.450$ and a K-S test statistic from $0.210$ to $0.460$. This suggests that the higher mass single stars, between $1.39\,M_\odot$ and $3.42\,M_\odot$, do not exhibit the same cumulative growth as lower mass stars. In this sample, the average mass is $1.36\,M_\odot$. The distributions of single and binary stars are similar up to approximately $5\,\mathrm{arcmin}$, beyond which single stars show less concentration compared to binary stars.

Given the current dynamical state of the cluster and the number of stars (presumably larger at birth), it is unlikely that there has been sufficient time for dynamical mass segregation. If the cluster initially formed in a roughly spherical distribution without significant substructure, this suggests that binaries originated closer to the cluster center than single stars. Alternatively, simulations by \citet{2007ApJ...655L..45M} indicate that this outcome could result from star formation in smaller clumps that either formed mass-segregated or had short relaxation times, allowing for quick mass segregation before merging into the current cluster. These simulations conclude that clumps with initial mass segregation largely preserve this characteristic during merging. For clumps without initial mass segregation, dynamic mass segregation can occur before merging if their initial relaxation times are short, and this segregation is inherited by the merged cluster. Additionally, initial mass segregation and gas expulsion play a critical role in star cluster evolution \citep{2008MNRAS.386.2047M}. Regardless of the specific mechanism, this suggests a primordial, or near-primordial, central concentration of binaries during the formation process of the cluster, which is congruent with other young open clusters \citep{2007AJ....134.1368C,2007MNRAS.381L..40D,2008AJ....135..173S}.

\subsection{Comparative analysis with cluster members and PMS study}
\subsubsection{HD 159176}
HD 159176, a double-lined spectroscopic binary composed of O7V type stars, is located in the projected center of NGC 6383 and is responsible for ionizing the surrounding HII region \citep{2016ApJ...832..211P}. The estimated age of HD 159176 is $2.30-2.80$\,Myr \citep{2010AA...511A..25R}. Previous analyses suggested that HD 159176 might predate the cluster itself and could have initiated star formation in the cluster's core and surrounding areas \citep{1978MNRAS.182..607F}.

In the study by \citet{2018AA...610A..30A}, it was hypothesized that if HD 159176 were a blue straggler, it would imply an older age for NGC 6383, ranging between $6.00$ and $10.0$\,Myr. However, our analysis indicates that HD 159176 is not a member of NGC 6383. The proper motion ($\mu_\alpha^* = 2.620 \pm 0.010$ mas yr$^{-1}$, $\mu_\delta = -0.7970 \pm 0.0090$ mas yr$^{-1}$) and parallax ($1.150 \pm 0.024$ mas) are outside the $3\sigma$ width of the cluster posterior distribution of $\mu_\delta$ and $\varpi$, whose distributions can be seen in Fig. \ref{fig:proper_motion} and Fig. \ref{fig:parallax_distance}.

Despite its central position, these discrepancies in proper motion and parallax confirm that HD 159176 is not gravitationally bound to NGC 6383. Therefore, the hypothesis that HD 159176 is a blue straggler influencing the cluster's age must be reconsidered. Our study supports the notion that NGC 6383 has an age of approximately 4 Myr, independent of the characteristics of HD159176.

\subsubsection{NGC 6383 22}
NGC 6383 22\footnote{Cataloged in Simbad as \textsc{Gaia DR3 4054615634716139264}}, is identified as a $\lambda$ Bootis star. These stars are chemically peculiar A-type stars with abundance anomalies attributed to the accretion of metal-poor material. In a spectroscopic survey of metal-weak and emission-line stars in the Southern Hemisphere, \citet{2020MNRAS.499.2701M} found that $\lambda$ Boo stars, including NGC 6383 22, exhibit excesses at longer wavelengths, suggestive of circumstellar discs. This characteristic aligns with other stars in NGC 6383 and those studied in \citet{2019MNRAS.484.5102K}. However, NGC 6383 22, despite matching the position and parallax of the cluster, does not belong to it because of its proper motion ($\mu_\alpha^* = 1.790 \pm 0.030$ mas yr$^{-1}$, $\mu_\delta = -2.710 \pm 0.024$ mas yr$^{-1}$), which is more than $3\sigma$ away from the mean of the posterior distribution of $\mu$. This result indicates that this star is likely passing near or through the cluster.

\subsubsection{PMS Study Using VPHAS+}

\citet{2019MNRAS.484.5102K} examined 55 classical T-Tauri stars (CTTS) in the star-forming region Sh 2-012 and NGC 6383, utilizing optical photometry and Gaia astrometry. They reported a median age of $2.80 \pm 1.60\,\mathrm{Myr}$ for these stars, with masses ranging from $0.3$ to $1.0\,M_\odot$. Notably, $94\%$ of CTTS with near-infrared cross-matches conform to the near-infrared T-Tauri locus, and all stars with mid-infrared photometry showed signs of accreting circumstellar discs concentrated around NGC 6383.

Out of the 55 CTTS kinematic members of Sh 2-012, only 15 are catalogued as members or potential members of the cluster in this study, with most having a PMS probability over $0.6$\footnote{Except for Gaia DR3 4054615467234669440, Gaia DR3 4054617116502154368, Gaia DR3 4054618761456141056, and Gaia DR3 4054826809671572096, which have probabilities of 16, 12, 55, and 45 percent, respectively.}. The difference in member selection between the studies is primarily due to the differences in the membership determination methods. It should be noted that \citet{2019MNRAS.484.5102K} determined the membership of NGC 6383 by modeling the distribution of proper motions with a double-peaked Gaussian. Thus, the difference between this approach and the one adopted in the present paper surely induced a difference between the samples of cluster members selected in the two works. Only one T-Tauri source\footnote{Catalogued as Gaia DR3 4054567805963940352} is likely a binary, with a $0.82$ probability of being a binary star according to our analysis.

\subsubsection{Comparison with published catalogs}
We compared our membership list with the catalogs from \citet{2020AA...640A...1C}, \citet{2021ApJ...923..129J}, \citet{2022ApJS..262....7H}, \citet{2024arXiv240305143H}, and the members listed in SIMBAD. From \citet{2020AA...640A...1C}, we identified 148 common sources, found 97 sources not cross-matched in their catalog, and 173 sources unique to our selection. On the other hand, we find 161 matches with the sample of \citet{2021ApJ...923..129J}, with 123 sources not present in our data, and 160 not listed in their catalog. The catalog of \citet{2022ApJS..262....7H} includes 90 common sources, 47 not found in our data, and 231 not in their catalog. \citet{2024arXiv240305143H} had 202 common sources, with 120 unique to their catalog, and 119 unique to our results. Using the SIMBAD database, we confirmed 168 common sources, with 163 unique to SIMBAD and 153 unique to our sample. It should be noted that all of these studies used Gaia data.

\section{Conclusion}\label{sec:conclusion}

This study presents an analysis of the young open cluster NGC 6383, employing Bayesian analysis and machine learning techniques to identify cluster members and determine key parameters. Using the HDBSCAN algorithm, we identified 254 probable cluster members with membership probabilities exceeding 60 percent.

Our findings indicate that the cluster has a mode age of $3.53^{+1.40}_{-1.00}~\mathrm{Myr}$, with a distance of $1.110 \pm 0.060\,\mathrm{kpc}$ derived from the parallax-based distance estimation. The core and tidal radii of the cluster were determined to be $1.950 \pm 0.190\,\mathrm{arcmin}$ and $40.4 \pm 14.3\,\mathrm{arcmin}$, respectively. We also observed mass segregation among binary stars, indicating that NGC 6383 is not fully relaxed.

The CMD analysis revealed a well-defined main sequence and a population of PMS stars, indicating an age spread of star formation within the cluster ranging from $1$ to $6$ Myr ago, suggesting ongoing star formation. The Sagitta neural network provided additional insights into the probabilities of PMS stars, their ages, and the extinction affecting cluster members.

Our analysis aligns with recent studies, offering updated parameters for NGC 6383. A significant result is the identification of HD 159176 as a non-member of NGC 6383. Despite its central position and previous assumptions about its influence on the cluster's formation, our analysis of proper motion and parallax data indicates that HD 159176 is not gravitationally bound to NGC 6383. This has important implications for determining the cluster’s age, now estimated independently of this source’s characteristics.

Additionally, we observed that binary stars are more centrally concentrated compared to single stars, suggesting primordial mass segregation. The calculated half-mass relaxation time indicates that NGC 6383 is not yet fully relaxed. The observed central concentration of binaries, even for those with segregation times longer than the cluster’s age, suggests these stars may have formed closer to the cluster center. This finding is consistent with other studies and simulations where initial mass segregation and gas expulsion play critical roles in star cluster evolution, which is in agreement with the studies of \citet{2007ApJ...655L..45M,2008MNRAS.386.2047M,2007AJ....134.1368C,2007MNRAS.381L..40D,2008AJ....135..173S}. Regardless of the specific mechanism, this suggests a primordial, or near-primordial, central concentration of binaries during the formation process of the cluster, which is consistent with other young open clusters in those studies and simulations.

Overall, this study demonstrates the effectiveness of combining advanced computational techniques with traditional astronomical methods to enhance our understanding of stellar clusters. The characterization of NGC 6383 provides a solid foundation for future investigations into its formation history and evolution. This study is a precursor for future work investigating mass segregation, pre-main sequence stars, age, and membership of new open cluster candidates in the Gaia era. The software used in this study is available on GitHub\footnote{\href{https://github.com/notluquis/COSMIC}{https://github.com/notluquis/COSMIC}}, and the catalog will be archived at the CDS.

\begin{acknowledgements}
We gratefully acknowledge support by the ANID BASAL project FB210003 and the SOCHIAS GEMINI project 32230014.

L.P. gratefully acknowledges the invaluable comments and guidance of Dr. Robert Mathieu in the preparation of this investigation.

This work has made use of data from the European Space Agency (ESA) mission {\it Gaia} (\url{https://www.cosmos.esa.int/gaia}), processed by the {\it Gaia} Data Processing and Analysis Consortium (DPAC, \url{https://www.cosmos.esa.int/web/gaia/dpac/consortium}). Funding for the DPAC has been provided by national institutions, in particular the institutions participating in the {\it Gaia} Multilateral Agreement. 

This publication makes use of data products from the Two Micron All Sky Survey, which is a joint project of the University of Massachusetts and the Infrared Processing and Analysis Center/California Institute of Technology, funded by the National Aeronautics and Space Administration and the National Science Foundation. This work made use of Astropy:\footnote{http://www.astropy.org} a community-developed core Python package and an ecosystem of tools and resources for astronomy \citep{astropy:2013, astropy:2018, astropy:2022}.
\end{acknowledgements}

\bibliographystyle{aa}
\bibliography{cites} 

\begin{appendix}

\section{HDBSCAN diagnostic and ASteCA posterior corner plots}

The present section describes the runs of HDBSCAN and presents the resulting plots, providing a comprehensive visualization of the clustering process. These figures illustrate the cluster size variation as a function of minimum cluster size, the hierarchical structure of the cluster system through a condensed cluster tree, and the posterior distributions of inferred parameters.

\begin{figure}
\centering
\resizebox{\hsize}{!}{\includegraphics{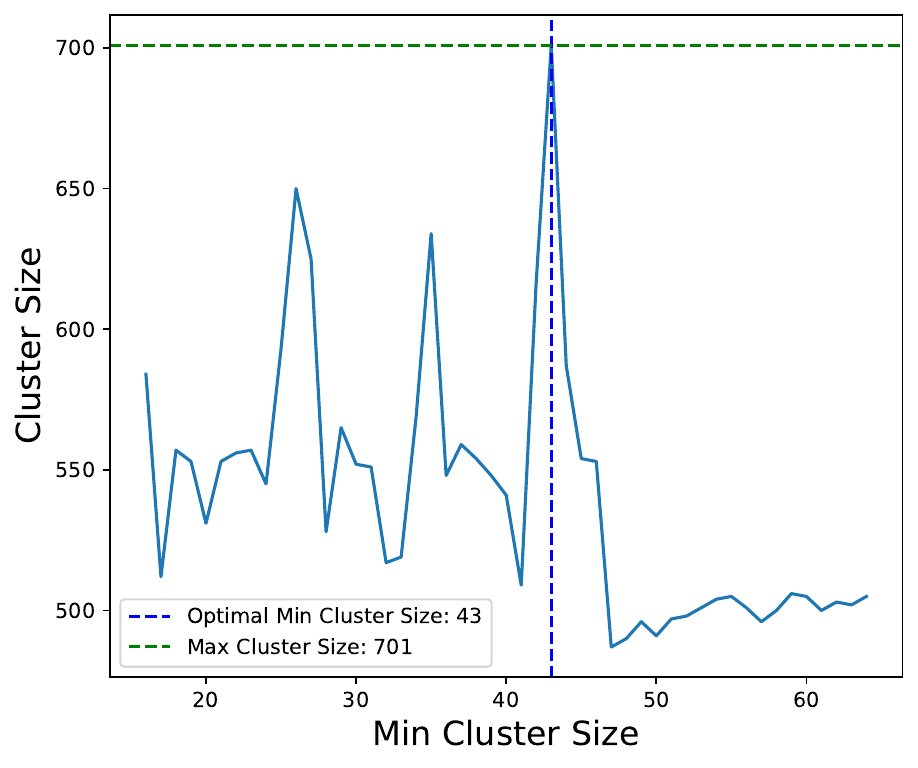}}
\caption{Cluster size variation as a function of minimum cluster size, determined through HDBSCAN clustering on the proper motion data of the sources. The blue line represents cluster sizes achieved at various minimum cluster sizes. The dashed blue line highlights the optimal minimum cluster size of 43, where the cluster size reaches a peak. The dashed green line indicates the maximum observed cluster size of 701.}
\label{fig:min_cluster_size}
\end{figure}

Figure \ref{fig:min_cluster_size} shows how the size of clusters varies with different minimum cluster size parameters in HDBSCAN. This plot helps visualizing the optimal minimum cluster size, marked by the dashed blue line, where the cluster size reaches its maximum efficiency. The dashed green line represents the maximum observed cluster size, providing a benchmark for comparison.

\begin{figure}
\resizebox{\hsize}{!}{\includegraphics{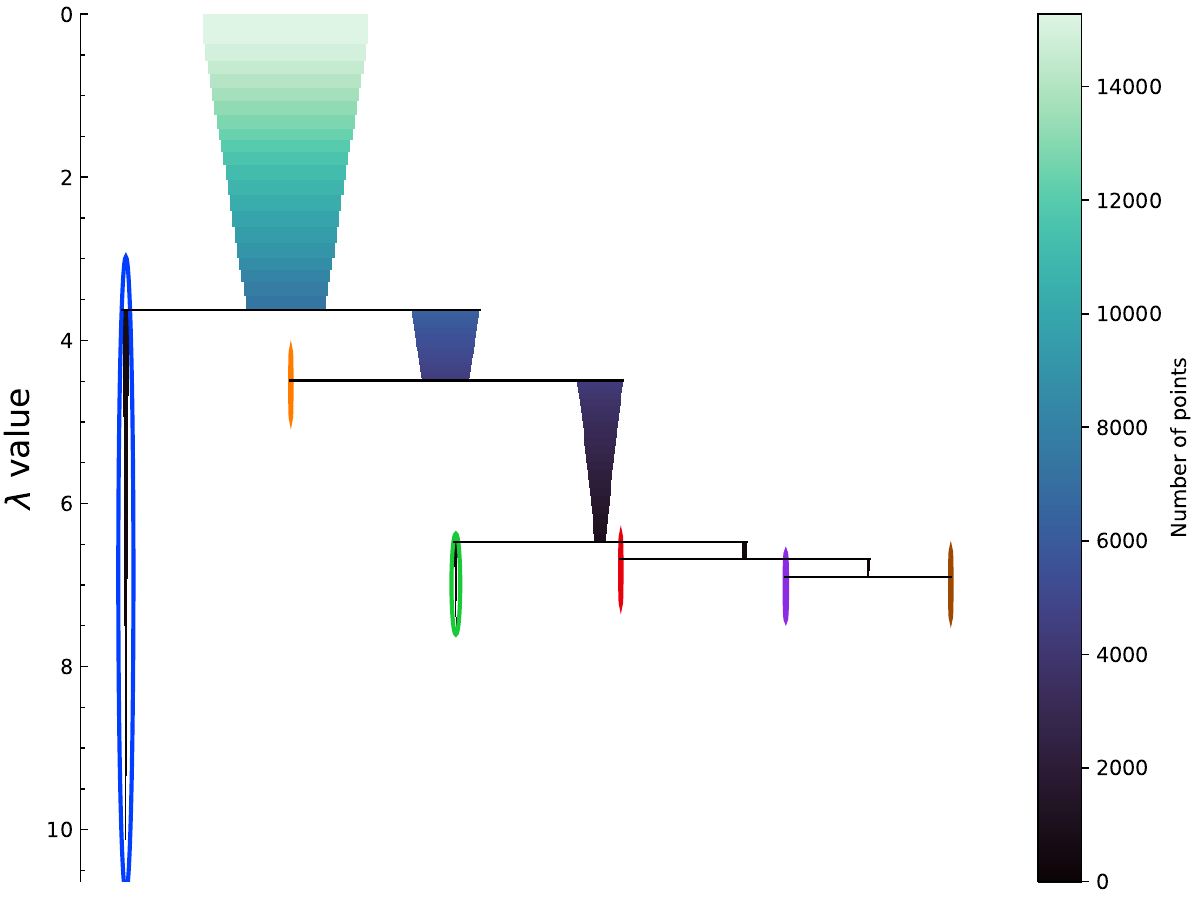}}
\caption{The condensed cluster tree exhibits the hierarchical structure of the cluster system, elucidating the proximity or potential merger of clusters as well as their degree of separation. The dendrogram representation displays the cluster hierarchy, wherein the width (and color) of each branch depicts the number of points in the cluster at that level. Specifically, the plot illustrates the NGC 6383 cluster sources on the left side, while the HDBSCAN-identified field sources are situated on the right side. The color bar corresponds to the number of sources at each level, while the $\lambda$ value corresponds to $\frac{1}{\texttt{distance}}$.}
\label{fig:condensed_cluster_tree}
\end{figure}

Figure \ref{fig:condensed_cluster_tree} presents the condensed cluster tree, which visually represents the hierarchical structure of the clusters. This dendrogram highlights the proximity or potential mergers of clusters and their degree of separation. The left side of the plot focuses on the NGC 6383 cluster sources, while the right side shows the HDBSCAN-identified field sources. The color bar indicates the number of sources at each hierarchical level, with $\lambda$ values representing the inverse of the distance.

\begin{figure*}
\resizebox{\hsize}{!}{\includegraphics{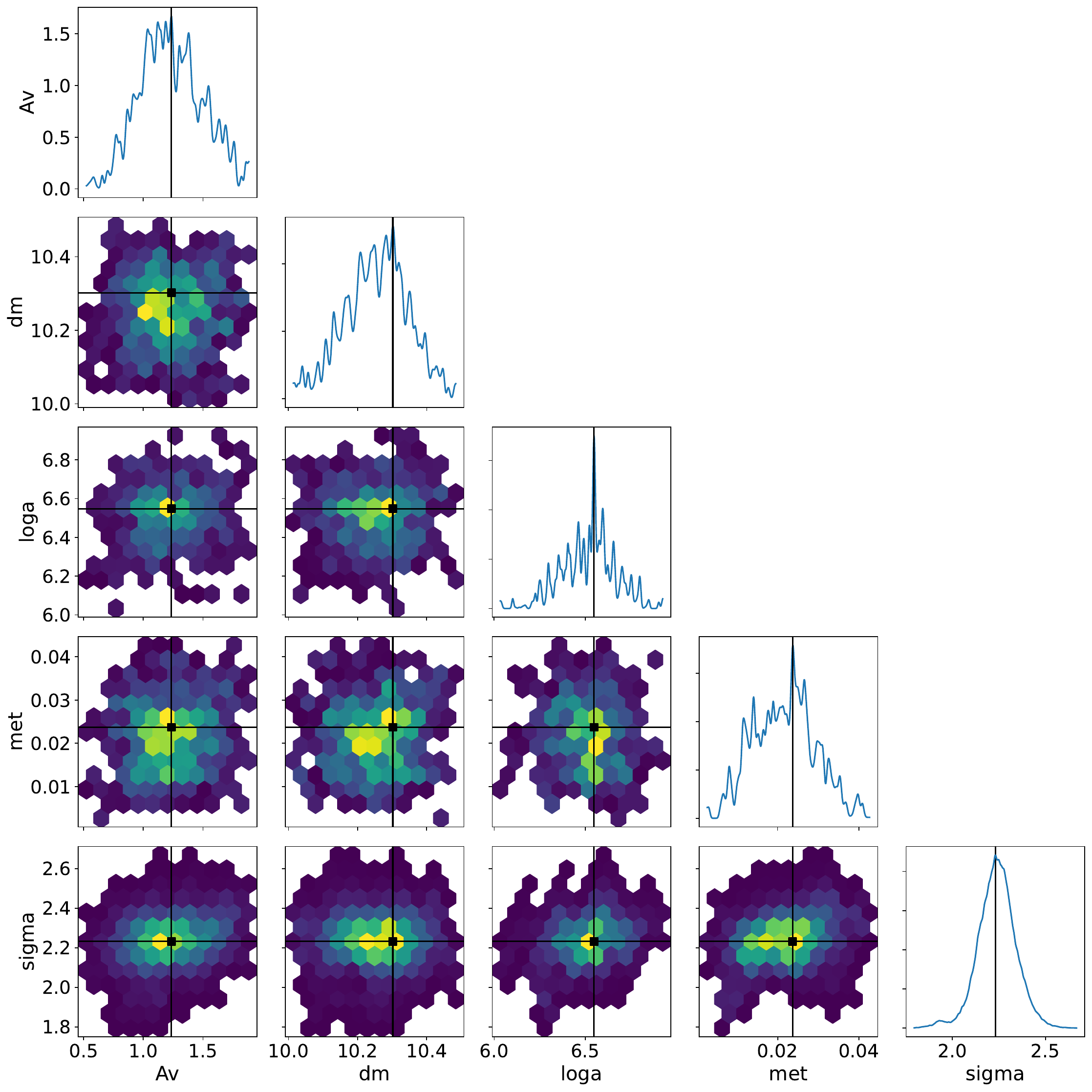}}
\caption{Posterior distributions of the parameters $A_v$ (visual extinction), $dm$ (distance modulus), $loga$ (logarithmic age), and $met$ (metallicity). The diagonals display the marginal distributions for each parameter, showing their individual probability densities. The off-diagonal hexbin plots illustrate the joint distributions, highlighting correlations between parameters. The black lines represent the mode of the distributions.}
\label{fig:plot_pair_trace}
\end{figure*}

Figure \ref{fig:plot_pair_trace} displays the posterior distributions of key parameters: visual extinction ($A_v$), distance modulus ($dm$), logarithmic age ($loga$), and metallicity ($met$). The diagonal plots show the marginal distributions for each parameter, providing insights into their individual probability densities. The off-diagonal hexbin plots illustrate the joint distributions, in which it is possible to see correlations between the parameters. The black lines indicate the mode of the distributions, highlighting the most probable values.

\end{appendix}
\end{document}